\documentclass[copyright,creativecommons]{eptcs}

\usepackage{tikzConf}

\usepackage[top=3cm,bottom=3cm,left=3cm,right=3cm]{geometry}
\usepackage[characters]{ltl}
\usepackage{listings}
\usepackage{verbatim}
\usepackage{url}
\usepackage{multirow}
\usepackage{xspace}
\usepackage{float}
\usepackage{xcolor}

\colorlet{codecolor}{blue!70!black}

\newcommand{\bools}{\ensuremath{\mathbb{B}}}
\newcommand{\nats}{\ensuremath{\mathbb{N}}}
\newcommand{\enums}{\ensuremath{\mathbb{E}}}
\newcommand{\buses}{\ensuremath{\mathbb{U}}}
\newcommand{\temporals}{\ensuremath{\mathbb{T}}}
\newcommand{\signals}{\ensuremath{\mathbb{S}}}
\newcommand{\pats}{\ensuremath{\mathbb{P}}}
\newcommand{\arbitrary}{\ensuremath{\mathbb{X}}}
\newcommand{\signalids}{\ensuremath{\Gamma_{\signals}}}
\newcommand{\busids}{\ensuremath{\Gamma_{\buses}}}
\newcommand{\enumids}{\ensuremath{\Gamma_{\enums}}}
\newcommand{\natids}{\ensuremath{\Gamma_{\nats}}}
\newcommand{\boolids}{\ensuremath{\Gamma_{\bools}}}
\newcommand{\temporalids}{\ensuremath{\Gamma_{\temporals}}}
\newcommand{\atypeids}{\ensuremath{\Gamma_{\mathcal{S}_{\arbitrary}}}}
\newcommand{\lstilde}{{\ensuremath{\color{codecolor} \raisebox{-9pt}{\scalebox{1.7}{\textasciitilde}}}}}
\newcommand{\tlsfsec}[1]{\ensuremath{\langle \text{\textit{#1}} \rangle}}
\newcommand{\tlsfid}[1]{\ensuremath{\langle \text{\textit{#1}} \rangle}}
\newcommand{\inputs}{\ensuremath{\mathcal{I}}}
\newcommand{\outputs}{\ensuremath{\mathcal{O}}}
\newcommand{\sep}{\ensuremath{\ \ | \ \ }}
\newcommand{\syntcomp}{SYNTCOMP\xspace}
\newcommand{\ltlf}{\ensuremath{\mathbf{LTL}_f}\xspace}
\usepackage{tikz}

\usetikzlibrary{calc}
\usetikzlibrary{automata}
\usetikzlibrary{backgrounds}
\usetikzlibrary{decorations.pathreplacing}

\renewcommand{\subsubsection}[1]{\medskip \noindent {\bf #1.}}

\newcommand{\src}[1]{\texttt{\lstinline!#1!}}
\newcommand{\secref}[1]{Sect.~\ref{#1}}

\newcommand{\oper}[1]{\texttt{\lstinline|#1|}}

\lstdefinestyle{TLSF}{
  belowcaptionskip=1\baselineskip,
  breaklines=true,
  frame=L,
  xleftmargin=\parindent,
  language=C,
  showstringspaces=false,
  basicstyle=\ttfamily\color{codecolor},
  keywordstyle=\ttfamily\color{codecolor},
  commentstyle=\ttfamily\color{codecolor},
  identifierstyle=\text\ttfamily\color{codecolor},
  stringstyle=\ttfamily\color{codecolor},
}

\lstdefinestyle{tlsf_keyword}{
  belowcaptionskip=1\baselineskip,
  breaklines=true,
  frame=L,
  xleftmargin=\parindent,
  language=C,
  showstringspaces=false,
  basicstyle=\ttfamily\color{codecolor},
  keywordstyle=\ttfamily\color{codecolor},
  commentstyle=\ttfamily\color{codecolor},
  identifierstyle=\text\ttfamily\color{codecolor},
  stringstyle=\ttfamily\color{codecolor},
}

\lstdefinestyle{tlsf_enumtyp}{
  belowcaptionskip=1\baselineskip,
  breaklines=true,
  frame=L,
  xleftmargin=\parindent,
  language=C,
  showstringspaces=false,
  basicstyle=\ttfamily\color{violet!90!black},
  keywordstyle=\ttfamily\color{violet!90!black},
  commentstyle=\ttfamily\color{violet!90!black},
  identifierstyle=\text\ttfamily\color{violet!90!black},
  stringstyle=\ttfamily\color{violet!90!black},
}

\lstdefinestyle{tlsf_enumval}{
  belowcaptionskip=1\baselineskip,
  breaklines=true,
  frame=L,
  xleftmargin=\parindent,
  language=C,
  showstringspaces=false,
  basicstyle=\ttfamily\color{red!60!black},
  keywordstyle=\ttfamily\color{red!60!black},
  commentstyle=\ttfamily\color{red!60!black},
  identifierstyle=\text\ttfamily\color{red!60!black},
  stringstyle=\ttfamily\color{red!60!black},
}

\lstdefinestyle{tlsf_semanti}{
  belowcaptionskip=1\baselineskip,
  breaklines=true,
  frame=L,
  xleftmargin=\parindent,
  language=C,
  showstringspaces=false,
  basicstyle=\ttfamily\color{gray!70!black},
  keywordstyle=\ttfamily\color{gray!70!black},
  commentstyle=\ttfamily\color{gray!70!black},
  identifierstyle=\text\ttfamily\color{gray!70!black},
  stringstyle=\ttfamily\color{gray!70!black},
}

\lstdefinestyle{tlsf_istring}{
  belowcaptionskip=1\baselineskip,
  breaklines=true,
  frame=L,
  xleftmargin=\parindent,
  language=C,
  showstringspaces=false,
  basicstyle=\ttfamily\color{red!30!orange!80!black},
  keywordstyle=\ttfamily\color{red!30!orange!80!black},
  commentstyle=\ttfamily\color{red!30!orange!80!black},
  identifierstyle=\text\ttfamily\color{red!30!orange!80!black},
  stringstyle=\ttfamily\color{red!30!orange!80!black},
}

\lstdefinestyle{tlsf_variabl}{
  belowcaptionskip=1\baselineskip,
  breaklines=true,
  frame=L,
  xleftmargin=\parindent,
  language=C,
  showstringspaces=false,
  basicstyle=\ttfamily\color{green!50!black},
  keywordstyle=\ttfamily\color{green!50!black},
  commentstyle=\ttfamily\color{green!50!black},
  identifierstyle=\text\ttfamily\color{green!50!black},
  stringstyle=\ttfamily\color{green!50!black},
}

\lstdefinestyle{tlsf_operato}{
  belowcaptionskip=1\baselineskip,
  breaklines=true,
  frame=L,
  xleftmargin=\parindent,
  language=C,
  showstringspaces=false,
  basicstyle=\ttfamily\color{cyan!50!black},
  keywordstyle=\ttfamily\color{cyan!50!black},
  commentstyle=\ttfamily\color{cyan!50!black},
  identifierstyle=\text\ttfamily\color{cyan!50!black},
  stringstyle=\ttfamily\color{cyan!50!black},
}

\lstdefinestyle{tlsf_default}{
  belowcaptionskip=1\baselineskip,
  breaklines=true,
  frame=L,
  xleftmargin=\parindent,
  language=C,
  showstringspaces=false,
  basicstyle=\ttfamily\color{black},
  keywordstyle=\ttfamily\color{black},
  commentstyle=\ttfamily\color{black},
  identifierstyle=\text\ttfamily\color{black},
  stringstyle=\ttfamily\color{black},
}

\lstdefinestyle{tlsf_comment}{
  belowcaptionskip=1\baselineskip,
  breaklines=true,
  frame=L,
  xleftmargin=\parindent,
  language=C,
  showstringspaces=false,
  basicstyle=\ttfamily\color{orange!60!black},
  keywordstyle=\ttfamily\color{orange!60!black},
  commentstyle=\ttfamily\color{orange!60!black},
  identifierstyle=\text\ttfamily\color{orange!60!black},
  stringstyle=\ttfamily\color{orange!60!black},
}

\DeclareMathAlphabet{\mathcal}{OMS}{cmsy}{m}{n}

\usepackage[capitalise,english,nameinlink]{cleveref} 

\providecommand{\thisvolume}[1]{this volume of {\sl Electronic
  Proceedings in Theoretical Computer Science}. Open Publishing Association}

\title{The Temporal Logic Synthesis Format TLSF v1.2}
\author{
Swen Jacobs
\institute{CISPA Helmholtz Center for Information Security\\ Saarbr\"ucken, Germany}
\email{jacobs@cispa.de}
\and
Guillermo A. P\'erez
\institute{University of Antwerp -- Flanders Make\\ Antwerp, Belgium}
\email{guillermo.perez@uantwerpen.be}
\and
Philipp Schlehuber-Caissier
\institute{EPITA Research Laboratory\\ Paris, France}
\email{philipp@lrde.epita.fr}
}
\def\titlerunning{TLSF v1.2}
\def\authorrunning{ } 
\begin{document}
\newcommand{\arbiter}[2]{
  \begin{scope}[shift={(#1,#2)}]
    \def\h{4}
    \def\w{5}

    \coordinate (ANEW) at (0.5*\w-\w/2,\h/2);
    \coordinate (ASIG) at (0.3*\w-\w/2,-\h/2);
    \coordinate (AREADY) at (0.7*\w-\w/2,-\h/2);
    \coordinate (AR0) at (-\w/2,0.75*\h-\h/2);
    \coordinate (AR1) at (-\w/2,0.6*\h-\h/2);
    \coordinate (ARd) at (-\w/2,0.475*\h-\h/2);
    \coordinate (ARn) at (-\w/2,0.3*\h-\h/2);
    \coordinate (AG0) at (\w/2,0.75*\h-\h/2);
    \coordinate (AG1) at (\w/2,0.6*\h-\h/2);
    \coordinate (AGd) at (\w/2,0.475*\h-\h/2);
    \coordinate (AGn) at (\w/2,0.3*\h-\h/2);
    \coordinate (AHGRANT) at (\w/2+1.5*\a,0);
    \coordinate (AHBUSREQ) at (-\w/2-1.5*\a,0);

    \draw[fill=cmodule] (-\w/2,-\h/2) rectangle (\w/2,\h/2);
    \node at (0,-0.05*\h) {\large\bf ARBITER};
    
    \path[thick]
    (ANEW) edge[<-] +(0,-\a)
    (ASIG) edge[<-] +(0,\a)
    (AREADY) edge[->] +(0,\a)
    (AR0) edge[->] +(\a,0)
    (AR1) edge[->] +(\a,0)
    (ARn) edge[->] +(\a,0)
    (AG0) edge[<-] +(-\a,0)
    (AG1) edge[<-] +(-\a,0)
    (AGn) edge[<-] +(-\a,0)
    ;
    
    \path
    (AG0) edge  +(\a/2,0)
    (AG1) edge  +(\a/2,0)
    (AGn) edge  +(\a/2,0)
    ($ (AG0) + (\a/2,0) $) edge ([yshift=4] AHGRANT)
    ($ (AG1) + (\a/2,0) $) edge ([yshift=2] AHGRANT)
    ($ (AGn) + (\a/2,0) $) edge ([yshift=-4] AHGRANT)
    ;

    \path
    (AR0) edge  +(-\a/2,0)
    (AR1) edge  +(-\a/2,0)
    (ARn) edge  +(-\a/2,0)
    ($ (AR0) + (-\a/2,0) $) edge ([yshift=4] AHBUSREQ)
    ($ (AR1) + (-\a/2,0) $) edge ([yshift=2] AHBUSREQ)
    ($ (ARn) + (-\a/2,0) $) edge ([yshift=-4] AHBUSREQ)
    ;

    \node at ($ (ANEW) + (0,-1.5*\a) $) {\small \textsc{decide}};
    \node at ($ (ASIG) + (0,1.5*\a) $) {\small \textsc{busreq}};
    \node at ($ (AREADY) + (0,1.5*\a) $) {\small \textsc{allready}};

    \node[anchor=west] at ($ (AR0) + (0.8*\a,0) $) {\small \textsc{hbusreq}$ _{0} $};        
    \node[anchor=west] at ($ (AR1) + (0.8*\a,0) $) {\small \textsc{hbusreq}$ _{1} $};        
    \node[anchor=west] at ($ (ARd) + (0.5*\a,0) $) {$ \vdots $};        
    \node[anchor=west] at ($ (ARn) + (0.8*\a,0) $) {\small \textsc{hbusreq}$ _{n-1} $};        

    \node[anchor=east] at ($ (AG0) + (-0.8*\a,0) $) {\small \textsc{hgrant}$ _{0} $};        
    \node[anchor=east] at ($ (AG1) + (-0.8*\a,0) $) {\small \textsc{hgrant}$ _{1} $};        
    \node[anchor=east] at ($ (AGd) + (-0.5*\a,0) $) {$ \vdots $};        
    \node[anchor=east] at ($ (AGn) + (-0.8*\a,0) $) {\small \textsc{hgrant}$ _{n-1} $};        
  \end{scope}
}

\newcommand{\lock}[2]{
  \begin{scope}[shift={(#1,#2)}]
    \def\h{4}
    \def\w{4.5}

    \coordinate (MSIG) at (0.5*\w-\w/2,\h/2);
    \coordinate (MUPD) at (0.5*\w-\w/2,-\h/2); 
    \coordinate (MX0) at (-\w/2,0.75*\h-\h/2);
    \coordinate (MX1) at (-\w/2,0.6*\h-\h/2);
    \coordinate (MXd) at (-\w/2,0.475*\h-\h/2);
    \coordinate (MXn) at (-\w/2,0.3*\h-\h/2);
    \coordinate (ME0) at (\w/2,0.75*\h-\h/2);
    \coordinate (ME1) at (\w/2,0.6*\h-\h/2);
    \coordinate (MEd) at (\w/2,0.475*\h-\h/2);
    \coordinate (MEn) at (\w/2,0.3*\h-\h/2);
    \coordinate (MHGRANT) at (\w/2+1.5*\a,0);
    \coordinate (MHLOCK) at (-\w/2-1.5*\a,0);

    \draw[fill=cmodule] (-\w/2,-\h/2) rectangle (\w/2,\h/2);
    \node at (0,-0.05*\h) {\large\bf LOCK};
    
    \path[thick]
    (MSIG) edge[<-] +(0,-\a)
    (MUPD) edge[->] +(0,\a)
    (MX0) edge[->] +(\a,0)
    (MX1) edge[->] +(\a,0)
    (MXn) edge[->] +(\a,0)
    (ME0) edge[->] +(-\a,0)
    (ME1) edge[->] +(-\a,0)
    (MEn) edge[->] +(-\a,0)
    ;

    \path
    (ME0) edge  +(\a/2,0)
    (ME1) edge  +(\a/2,0)
    (MEn) edge  +(\a/2,0)
    ($ (ME0) + (\a/2,0) $) edge ([yshift=4] MHGRANT)
    ($ (ME1) + (\a/2,0) $) edge ([yshift=2] MHGRANT)
    ($ (MEn) + (\a/2,0) $) edge ([yshift=-4] MHGRANT)
    ;

    \path
    (MX0) edge  +(-\a/2,0)
    (MX1) edge  +(-\a/2,0)
    (MXn) edge  +(-\a/2,0)
    ($ (MX0) + (-\a/2,0) $) edge ([yshift=4] MHLOCK)
    ($ (MX1) + (-\a/2,0) $) edge ([yshift=2] MHLOCK)
    ($ (MXn) + (-\a/2,0) $) edge ([yshift=-4] MHLOCK)
    ;

    \node at ($ (MSIG) + (0,-1.5*\a) $) {\small \textsc{locked}};
    \node at ($ (MUPD) + (0,1.5*\a) $) {\small \textsc{decide}};

    \node[anchor=west] at ($ (MX0) + (0.8*\a,0) $) {\small \textsc{hlock}$ _{0} $};        
    \node[anchor=west] at ($ (MX1) + (0.8*\a,0) $) {\small \textsc{hlock}$ _{1} $};        
    \node[anchor=west] at ($ (MXd) + (0.5*\a,0) $) {$ \vdots $};        
    \node[anchor=west] at ($ (MXn) + (0.8*\a,0) $) {\small \textsc{hlock}$ _{n-1} $};        

    \node[anchor=east] at ($ (ME0) + (-0.8*\a,0) $) {\small \textsc{hgrant}$ _{0} $};        
    \node[anchor=east] at ($ (ME1) + (-0.8*\a,0) $) {\small \textsc{hgrant}$ _{1} $};        
    \node[anchor=east] at ($ (MEd) + (-0.5*\a,0) $) {$ \vdots $};        
    \node[anchor=east] at ($ (MEn) + (-0.8*\a,0) $) {\small \textsc{hgrant}$ _{n-1} $};        
  \end{scope}
}

\newcommand{\decode}[2]{
  \begin{scope}[shift={(#1,#2)}]
    \def\h{4}
    \def\w{3.6}

    \coordinate (INCR) at (\w/2,0.75*\h-\h/2);
    \coordinate (BURST4) at (\w/2,0.5*\h-\h/2);
    \coordinate (SINGLE) at (\w/2,0.25*\h-\h/2);
    \coordinate (HBURST) at (-\w/2,0.35*\h-\h/2);

    \draw[fill=cmodule] (-\w/2,-\h/2) rectangle (\w/2,\h/2);
    \node at (-0.15*\w,0.25*\h) {\large\bf DECODE};
    
    \path[thick]
    (INCR) edge[<-] +(-\a,0)
    (BURST4) edge[<-] +(-\a,0)
    (SINGLE) edge[<-] +(-\a,0)
    (HBURST) edge +(1.4*\a,0)
    ([yshift=2] HBURST) edge +(0.9*\a,0)
    ([yshift=4] HBURST) edge +(0.7*\a,0)
    ([yshift=-2] HBURST) edge +(0.9*\a,0)
    ([yshift=-4] HBURST) edge +(0.7*\a,0)
    ($ (HBURST) + (0.6*\a,-0.3) $) edge[bend left=40,ultra thick,cap=round] ($ (HBURST) + (1.4*\a,0) $)
    ($ (HBURST) + (0.6*\a,0.3) $) edge[bend right=40,ultra thick,cap=round] ($ (HBURST) + (1.4*\a,0) $)
    ;

    \node[anchor=east] at ($ (INCR) + (-0.8*\a,0) $) {\small \textsc{incr}};
    \node[anchor=east] at ($ (BURST4) + (-0.8*\a,0) $) {\small \textsc{burst}\scalebox{0.85}{4}};
    \node[anchor=east] at ($ (SINGLE) + (-0.8*\a,0) $) {\small \textsc{single}};
    \node[anchor=west] at ($ (HBURST) + (1.4*\a,0) $) {\small \textsc{hburst}};
  \end{scope}
}

\newcommand{\encode}[2]{
  \begin{scope}[shift={(#1,#2)}]
    \def\h{4}
    \def\w{4}

    \coordinate (HMASTER) at (-\w/2,0.65*\h-\h/2);
    \coordinate (HMASTERS) at ($ (HMASTER) + (\w - 1.5*\a,0) $);
    \coordinate (EMASTER) at ($ (HMASTER) + (\w,0) $);
    \coordinate (EHREADY) at (-\w/2,0.65*\h-\h/2);
    \coordinate (EG0) at (0.25*\w-\w/2,-\h/2);
    \coordinate (EG1) at (0.4*\w-\w/2,-\h/2);
    \coordinate (EGd) at (0.525*\w-\w/2,-\h/2);
    \coordinate (EGn) at (0.7*\w-\w/2,-\h/2);
    \coordinate (DHGRANT) at (0,-\h/2-1.5*\a);

    \draw[fill=cmodule] (-\w/2,-\h/2) rectangle (\w/2,\h/2);
    \node at (0,0.35*\h) {\large\bf ENCODE};

    \path
    (EG0) edge +(0,-\a/2)
    (EG1) edge +(0,-\a/2)
    (EGn) edge +(0,-\a/2)
    ($ (EG0) - (0,\a/2) $) edge ([xshift=-4] DHGRANT)
    ($ (EG1) - (0,\a/2) $) edge ([xshift=-2] DHGRANT)
    ($ (EGn) - (0,\a/2) $) edge ([xshift=4] DHGRANT)
    ;      
    
    \path[thick]
    (HMASTERS) edge +(1.4*\a,0)
    (EHREADY) edge[->] +(\a,0)
    (EG0) edge[->] +(0,\a)
    (EG1) edge[->] +(0,\a)
    (EGn) edge[->] +(0,\a)
    ([yshift=2] HMASTERS) edge +(0.9*\a,0)
    ([yshift=4] HMASTERS) edge +(0.7*\a,0)
    ([yshift=-2] HMASTERS) edge +(0.9*\a,0)
    ([yshift=-4] HMASTERS) edge +(0.7*\a,0)
    ($ (HMASTERS) + (0.6*\a,-0.3) $) edge[bend left=40,ultra thick,cap=round] ($ (HMASTERS) + (1.4*\a,0) $)
    ($ (HMASTERS) + (0.6*\a,0.3) $) edge[bend right=40,ultra thick,cap=round] ($ (HMASTERS) + (1.4*\a,0) $)
    ;

    \node[anchor=east] at (HMASTERS) {\small \textsc{hmaster}};
    \node[anchor=west] at ($ (EHREADY) + (\a,0) $) {\small \textsc{hready}};
    \node[anchor=west,rotate=90] at ($ (EG0) + (0,0.8*\a) $) {\small \textsc{hgrant}$ _{0} $};
    \node[anchor=west,rotate=90] at ($ (EG1) + (0,0.8*\a) $) {\small \textsc{hgrant}$ _{1} $};
    \node[anchor=west,rotate=90] at ($ (EGd) + (0,0.5*\a) $) {\small $ \vdots $};
    \node[anchor=west,rotate=90] at ($ (EGn) + (0,0.8*\a) $) {\small \textsc{hgrant}$ _{n-1} $};
  \end{scope}
}

\newcommand{\shift}[2]{
  \begin{scope}[shift={(#1,#2)}]
    \def\h{2}
    \def\w{4}

    \coordinate (HMASTLOCK) at (\w/2,0.35*\h-\h/2);
    \coordinate (FHREADY) at (-\w/2,0.2*\h-\h/2); 
    \coordinate (SLOCKED) at (-\w/2,0.5*\h-\h/2); 

    \draw[fill=cmodule] (-\w/2,-\h/2) rectangle (\w/2,\h/2);
    \node at (0,0.28*\h) {\large\bf SHIFT};
    
    \path[thick]
    (HMASTLOCK) edge[<-] +(-\a,0)
    (FHREADY) edge[->] +(\a,0)
    (SLOCKED) edge[->] +(\a,0)
    ;

    \node[anchor=east] at ($ (HMASTLOCK) + (-\a,0) $) {\small \textsc{hmastlock}};
    \node[anchor=west] at ($ (FHREADY) + (\a,0) $) {\small \textsc{hready}};
    \node[anchor=west] at ($ (SLOCKED) + (\a,0) $) {\small \textsc{locked}};
  \end{scope}
}

\newcommand{\tincr}[2]{
  \begin{scope}[shift={(#1,#2)}]
    \def\h{4}
    \def\w{4.5}

    \coordinate (IREADY) at (0.5*\w-\w/2,\h/2);
    \coordinate (IHREADY) at (0.2*\w-\w/2,-\h/2);
    \coordinate (ILOCK) at (0.4*\w-\w/2,-\h/2);
    \coordinate (IDECIDE) at (0.6*\w-\w/2,-\h/2);
    \coordinate (ISIG) at (0.8*\w-\w/2,-\h/2);
    \coordinate (IBUSREQ) at (-\w/2,0.7*\h-\h/2);

    \draw[fill=cmodule] (-\w/2,-\h/2) rectangle (\w/2,\h/2);
    \node at (0.1*\w,0.1*\h) {\large\bf TINCR};
    
    \path[thick]
    (IREADY) edge[<-] +(0,-\a)      
    (IHREADY) edge[->] +(0,\a)
    (ILOCK) edge[->] +(0,\a)
    (IDECIDE) edge[->] +(0,\a)
    (ISIG) edge[->] +(0,\a)
    (IBUSREQ) edge[->] +(\a,0)
    ;

    \node at ($ (IREADY) + (0,-1.5*\a) $) {\small \textsc{ready}$ _{1} $};
    \node[rotate=90,anchor=west] at ($ (IHREADY) + (0,\a) $) {\small \textsc{hready}};
    \node[rotate=90,anchor=west] at ($ (ILOCK) + (0,\a) $) {\small \textsc{locked}};
    \node[rotate=90,anchor=west] at ($ (IDECIDE) + (0,\a) $) {\small \textsc{decide}};
    \node[rotate=90,anchor=west] at ($ (ISIG) + (0,\a) $) {\small \textsc{incr}};
    \node[anchor=west] at ($ (IBUSREQ) + (\a,0) $) {\small \textsc{busreq}};
  \end{scope}
}

\newcommand{\tburstfour}[2]{
  \begin{scope}[shift={(#1,#2)}]
    \def\h{4}
    \def\w{4.5}

    \coordinate (BREADY) at (0.5*\w-\w/2,\h/2);
    \coordinate (BHREADY) at (0.2*\w-\w/2,-\h/2);
    \coordinate (BLOCK) at (0.4*\w-\w/2,-\h/2);
    \coordinate (BDECIDE) at (0.6*\w-\w/2,-\h/2);
    \coordinate (BSIG) at (0.8*\w-\w/2,-\h/2);

    \draw[fill=cmodule] (-\w/2,-\h/2) rectangle (\w/2,\h/2);
    \node at (0,0.1*\h) {\large\bf TBURST4};
    
    \path[thick]
    (BREADY) edge[<-] +(0,-\a)      
    (BHREADY) edge[->] +(0,\a)
    (BLOCK) edge[->] +(0,\a)
    (BDECIDE) edge[->] +(0,\a)
    (BSIG) edge[->] +(0,\a)
    ;

    \node at ($ (BREADY) + (0,-1.5*\a) $) {\small \textsc{ready}$ _{2} $};
    \node[rotate=90,anchor=west] at ($ (BHREADY) + (0,\a) $) {\small \textsc{hready}};
    \node[rotate=90,anchor=west] at ($ (BLOCK) + (0,\a) $) {\small \textsc{locked}};
    \node[rotate=90,anchor=west] at ($ (BDECIDE) + (0,\a) $) {\small \textsc{decide}};
    \node[rotate=90,anchor=west] at ($ (BSIG) + (0,\a) $) {\small \textsc{burst}\scalebox{0.85}{4}};
  \end{scope}
}

\newcommand{\tsingle}[2]{
  \begin{scope}[shift={(#1,#2)}]
    \def\h{4}
    \def\w{4}

    \coordinate (SREADY) at (0.5*\w-\w/2,\h/2);
    \coordinate (SHREADY) at (0.2*\w-\w/2,-\h/2);
    \coordinate (SLOCK) at (0.4*\w-\w/2,-\h/2);
    \coordinate (SDECIDE) at (0.6*\w-\w/2,-\h/2);
    \coordinate (SSIG) at (0.8*\w-\w/2,-\h/2);

    \draw[fill=cmodule] (-\w/2,-\h/2) rectangle (\w/2,\h/2);
    \node at (0,0.1*\h) {\large\bf TSINGLE};
    
    \path[thick]
    (SREADY) edge[<-] +(0,-\a)      
    (SHREADY) edge[->] +(0,\a)
    (SLOCK) edge[->] +(0,\a)
    (SDECIDE) edge[->] +(0,\a)
    (SSIG) edge[->] +(0,\a)
    ;

    \node at ($ (SREADY) + (0,-1.5*\a) $) {\small \textsc{ready}$ _{3} $};
    \node[rotate=90,anchor=west] at ($ (SHREADY) + (0,\a) $) {\small \textsc{hready}};
    \node[rotate=90,anchor=west] at ($ (SLOCK) + (0,\a) $) {\small \textsc{locked}};
    \node[rotate=90,anchor=west] at ($ (SDECIDE) + (0,\a) $) {\small \textsc{decide}};
    \node[rotate=90,anchor=west] at ($ (SSIG) + (0,\a) $) {\small \textsc{single}};
  \end{scope}
}

\newcommand{\mand}[2]{
  \begin{scope}[shift={(#1,#2)}]
    \def\h{1.2}
    \def\w{2}

    \coordinate (R1) at (-0.3,-\h/2);
    \coordinate (R2) at (0,-\h/2);
    \coordinate (R3) at (0.3,-\h/2);
    \coordinate (AA) at (0,\h/2);

    \draw[fill=cmodule] (-\w/2,-\h/2) rectangle (\w/2,\h/2);
    \node at (0,0) {\large\bf AND};

    \path[thick]
    (R1) edge[->] +(0,\a)      
    (R2) edge[->] +(0,\a)      
    (R3) edge[->] +(0,\a)      
    (AA) edge[<-] +(0,-\a)      
    ;
  \end{scope}
}

\newsavebox{\architecture}
\begin{lrbox}{\architecture}
  \begin{tikzpicture}
    \def\a{0.3}
    \definecolor{cmodule}{gray}{0.9}
    \def\dd{0.7}

    \def\fleft{-9.5}
    \def\fright{7.6}
    \def\ftop{16.9}
    \def\fbot{-8.5}

    \def\cx{1.5}
    \def\cy{5.3}

    \arbiter{-4.8}{8}
    \lock{-4.8}{13.8}
    \decode{-7.1}{-6}
    \encode{5}{11.5}
    \shift{5}{15.4}
    \tincr{-6}{0}
    \tburstfour{-1}{0}
    \tsingle{3.75}{0}
    \mand{-1}{4}

    \draw (ANEW) -- (MUPD);
    \draw (ASIG) |- (\fleft+0.6,\cy) |- (IBUSREQ);
    \draw (INCR) -| (ISIG);
    \draw (BURST4) -| (BSIG);
    \draw (SINGLE) -| (SSIG);
    \draw (SHREADY) -- +(0,-1.3) node[fill,circle,inner sep=1pt] {} -| ($ (SREADY) + (3.3,2.3) $);
    \draw (SDECIDE) -- +(0,-0.7) node[fill,circle,inner sep=1pt] {}-| ($ (SREADY) + (2.7,1.7) $);
    \draw (SLOCK) -- +(0,-1)  node[fill,circle,inner sep=1pt] {} -| ($ (SREADY) + (3,2) $);
    \draw ($ (SREADY) + (3,2) $) -| (\cx,\ftop-0.5); 
    \draw ($ (SREADY) + (2.7,1.7) $) -| (\cx-0.3,5); 
    \draw ($ (SREADY) + (3.3,2.3) $) -| (\cx+0.3,5); 
    \draw (FHREADY) -| (\cx+0.3,5); 
    \draw (MSIG) |- (\cx,\ftop-0.5); 
    \draw (\cx-0.3,5) |- ($ (ANEW) + (0,0.9) $) node[fill,circle,inner sep=1pt] {};
    \draw (SLOCKED) -- (SLOCKED -| \cx,0) node[fill,circle,inner sep=1pt] {};
    \draw (EHREADY) -- (EHREADY -| \cx+0.3,0) node[fill,circle,inner sep=1pt] {};
    \draw (IDECIDE) -- +(0,-0.7) -- +(9.65,-0.7);
    \draw (BDECIDE) -- +(0,-0.7) node[fill,circle,inner sep=1pt] {};
    \draw (ILOCK) -- +(0,-1) -- +(9.8,-1);
    \draw (BLOCK) -- +(0,-1) node[fill,circle,inner sep=1pt] {};
    \draw (IHREADY) -- +(0,-1.3)  node[fill,circle,inner sep=1pt] {} -- +(9.95,-1.3); 
    \draw (BHREADY) -- +(0,-1.3) node[fill,circle,inner sep=1pt] {};

    \foreach \i in {-4,-2,0,2,4} 
      \draw ([xshift=\i] DHGRANT) -- ([xshift=\i,yshift=\i] AHGRANT -| DHGRANT)  node[fill,circle,inner sep=0.6pt] {};

    \foreach \i in {-4,-2,0,2,4} 
      \draw ([yshift=\i] MHGRANT) -| ($ ([xshift=\i,yshift=\i] AHGRANT) + (1.5,0) $) node[fill,circle,inner sep=0.6pt] {};

    \draw (BREADY) -- (R2); 
    \draw (IREADY) -- +(0,0.7) -| (R1);
    \draw (SREADY) -- +(0,0.7) -| (R3);

    \draw (AA) -- (AA |- 0,\cy) -| (AREADY);

    \node[above,yshift=5,xshift=6] at (AHGRANT -| \fright+\dd,0) {\small \textsc{HGRANT}};
    \foreach \i in {-4,-2,0,2,4} 
      \draw ([yshift=\i] AHGRANT) -- ([yshift=\i] AHGRANT -| \fright+\dd,0);

    \node[above,yshift=5,xshift=-3] at (MHLOCK -| \fleft-\dd,0) {\small \textsc{HLOCK}};
    \foreach \i in {-4,-2,0,2,4} 
      \draw ([yshift=\i] MHLOCK) -- ([yshift=\i] MHLOCK -| \fleft-\dd,0);

    \node[above,yshift=5,xshift=-9] at (AHBUSREQ -| \fleft-\dd,0) {\small \textsc{HBUSREQ}};
    \foreach \i in {-4,-2,0,2,4} 
      \draw ([yshift=\i] AHBUSREQ) -- ([yshift=\i] AHBUSREQ -| \fleft-\dd,0);

    \draw (HMASTLOCK) -- (HMASTLOCK -| \fright+\dd,0) node[above,xshift=16,yshift=1] {\small \textsc{HMASTLOCK}};

    \node[above,yshift=5,xshift=9.5] at (HMASTER -| \fright+\dd,0) {\small \textsc{HMASTER}};
    \foreach \i in {-4,-2,0,2,4} 
      \draw ([yshift=\i] EMASTER) -- ([yshift=\i] EMASTER -| \fright+\dd,0);

    \node[above,yshift=5,xshift=-5] at (HBURST -| \fleft-\dd,0) {\small \textsc{HBURST}};
    \foreach \i in {-4,-2,0,2,4} 
      \draw ([yshift=\i] HBURST) -- ([yshift=\i] HBURST -| \fleft-\dd,0);

    \coordinate (P1) at ($ (IHREADY) + (0,-1.3) $);
    \draw (P1) -- (P1 -| \fleft-\dd,0) node[above,xshift=-5,yshift=1] {\small \textsc{HREADY}};

    \draw[ultra thick] (\fleft,\fbot) rectangle (\fright,\ftop);
  \end{tikzpicture}
\end{lrbox}

\newsavebox{\cPArbiter}
\begin{lrbox}{\cPArbiter}
\begin{minipage}{0.7\textwidth}
\scriptsize
\noindent
\lstinline[style=tlsf_keyword]~INFO~%
\lstinline[style=tlsf_default]~ {~\\%
\lstinline[style=tlsf_keyword]~  TITLE~%
\lstinline[style=tlsf_default]~:       ~%
\lstinline[style=tlsf_istring]~"A Parameterized Arbiter"~\\%
\lstinline[style=tlsf_keyword]~  DESCRIPTION~%
\lstinline[style=tlsf_default]~: ~%
\lstinline[style=tlsf_istring]~"An arbiter, parameterized in the number of clients"~\\%
\lstinline[style=tlsf_keyword]~  SEMANTICS~%
\lstinline[style=tlsf_default]~:   ~%
\lstinline[style=tlsf_semanti]~Mealy~\\%
\lstinline[style=tlsf_keyword]~  TARGET~%
\lstinline[style=tlsf_default]~:      ~%
\lstinline[style=tlsf_semanti]~Mealy~\\%
\lstinline[style=tlsf_default]~}~\\%
\lstinline[style=tlsf_keyword]~GLOBAL ~%
\lstinline[style=tlsf_default]~{~\\%
\lstinline[style=tlsf_keyword]~  PARAMETERS ~%
\lstinline[style=tlsf_default]~{~\\%
\lstinline[style=tlsf_comment]~    // two clients~\\%
\lstinline[style=tlsf_variabl]~    n~%
\lstinline[style=tlsf_default]~ = 2;~\\%
\lstinline[style=tlsf_default]~  }~\\%
\lstinline[style=tlsf_keyword]~  DEFINITIONS~%
\lstinline[style=tlsf_default]~ {~\\%
\lstinline[style=tlsf_comment]~    // mutual exclusion~\\%
\lstinline[style=tlsf_variabl]~    mutual~%
\lstinline[style=tlsf_default]~(b) =~\\%
\lstinline[style=tlsf_operato]~      ||~%
\lstinline[style=tlsf_default]~[i ~%
\lstinline[style=tlsf_operato]~IN~%
\lstinline[style=tlsf_default]~ {0, 1 .. (~%
\lstinline[style=tlsf_operato]~SIZEOF ~%
\lstinline[style=tlsf_default]~b)~%
\lstinline[style=tlsf_operato]~ - ~%
\lstinline[style=tlsf_default]~1}]~\\%
\lstinline[style=tlsf_operato]~        &&~%
\lstinline[style=tlsf_default]~[j ~%
\lstinline[style=tlsf_operato]~IN~%
\lstinline[style=tlsf_default]~ {0, 1 .. (~%
\lstinline[style=tlsf_operato]~SIZEOF ~%
\lstinline[style=tlsf_default]~b)~%
\lstinline[style=tlsf_operato]~ - ~%
\lstinline[style=tlsf_default]~1} ~%
\lstinline[style=tlsf_operato]~(\)~%
\lstinline[style=tlsf_default]~ {i}]~\\%
\lstinline[style=tlsf_operato]~          !~%
\lstinline[style=tlsf_default]~(b[i] ~%
\lstinline[style=tlsf_operato]~&&~%
\lstinline[style=tlsf_default]~ b[j]);~\\%
\lstinline[style=tlsf_comment]~    // the Request-Response condition~\\%
\lstinline[style=tlsf_variabl]~    reqres~%
\lstinline[style=tlsf_default]~(req, res) =~\\%
\lstinline[style=tlsf_operato]~      G~%
\lstinline[style=tlsf_default]~ (req ~%
\lstinline[style=tlsf_operato]~-> F~%
\lstinline[style=tlsf_default]~ res);~\\%
\lstinline[style=tlsf_default]~  }~\\%
\lstinline[style=tlsf_default]~}~\\%
\lstinline[style=tlsf_comment]~/* ensure mutual exclusion on the output bus and guarantee~\\%
\lstinline[style=tlsf_comment]~   that each request on the input bus is eventually granted */~\\%
\lstinline[style=tlsf_keyword]~MAIN~%
\lstinline[style=tlsf_default]~ {~\\%
\lstinline[style=tlsf_keyword]~  INPUTS~%
\lstinline[style=tlsf_default]~ {~\\%
\lstinline[style=tlsf_variabl]~    r~%
\lstinline[style=tlsf_default]~[n];~\\%
\lstinline[style=tlsf_default]~  }~\\%
\lstinline[style=tlsf_keyword]~  OUTPUTS~%
\lstinline[style=tlsf_default]~ {~\\%
\lstinline[style=tlsf_variabl]~    g~%
\lstinline[style=tlsf_default]~[n];~\\%
\lstinline[style=tlsf_default]~  }~\\%
\lstinline[style=tlsf_keyword]~  ASSERT~%
\lstinline[style=tlsf_default]~ {~\\%
\lstinline[style=tlsf_default]~    mutual(g);~\\%
\lstinline[style=tlsf_default]~  }~\\%
\lstinline[style=tlsf_keyword]~  GUARANTEE~%
\lstinline[style=tlsf_default]~ {~\\%
\lstinline[style=tlsf_operato]~    &&~%
\lstinline[style=tlsf_default]~[0 ~%
\lstinline[style=tlsf_operato]~<=~%
\lstinline[style=tlsf_default]~ i ~%
\lstinline[style=tlsf_operato]~<~%
\lstinline[style=tlsf_default]~ n]~\\%
\lstinline[style=tlsf_default]~      reqres(r[i], g[i]);~\\%
\lstinline[style=tlsf_default]~  }~\\
\lstinline[style=tlsf_default]~}~
\end{minipage}
\end{lrbox}

\newsavebox{\cDecodeA}
\begin{lrbox}{\cDecodeA}
\begin{minipage}{0.4\textwidth}
\scriptsize
\lstinline[style=tlsf_keyword]~INFO~%
\lstinline[style=tlsf_default]~ {~\\%
\lstinline[style=tlsf_keyword]~  TITLE~%
\lstinline[style=tlsf_default]~:       ~%
\lstinline[style=tlsf_istring]~"AMBA AHB Arbiter"~\\%
\lstinline[style=tlsf_keyword]~  DESCRIPTION~%
\lstinline[style=tlsf_default]~: ~%
\lstinline[style=tlsf_istring]~"Component: Decode"~\\%
\lstinline[style=tlsf_keyword]~  SEMANTICS~%
\lstinline[style=tlsf_default]~:   ~%
\lstinline[style=tlsf_semanti]~Mealy~\\%
\lstinline[style=tlsf_keyword]~  TARGET~%
\lstinline[style=tlsf_default]~:      ~%
\lstinline[style=tlsf_semanti]~Mealy~\\%
\lstinline[style=tlsf_default]~}~\\%
\lstinline[style=tlsf_default]~~\\%
\lstinline[style=tlsf_keyword]~GLOBAL ~%
\lstinline[style=tlsf_default]~{~\\%
\lstinline[style=tlsf_keyword]~  DEFINITIONS~%
\lstinline[style=tlsf_default]~ {~\\%
\lstinline[style=tlsf_keyword]~    enum~%
\lstinline[style=tlsf_enumtyp]~ hburst~%
\lstinline[style=tlsf_default]~ =~\\%
\lstinline[style=tlsf_enumval]~      Single~%
\lstinline[style=tlsf_default]~: 00~\\%
\lstinline[style=tlsf_enumval]~      Burst4~%
\lstinline[style=tlsf_default]~: 10~\\%
\lstinline[style=tlsf_enumval]~      Incr~%
\lstinline[style=tlsf_default]~:   01~\\%
\lstinline[style=tlsf_default]~  }~\\%
\lstinline[style=tlsf_default]~}~\\%
\lstinline[style=tlsf_default]~~
\end{minipage}
\end{lrbox}

\newsavebox{\cDecodeB}
\begin{lrbox}{\cDecodeB}
\begin{minipage}{0.528\textwidth}
\scriptsize
\lstinline[style=tlsf_keyword]~MAIN~%
\lstinline[style=tlsf_default]~ {~\\%
\lstinline[style=tlsf_keyword]~  INPUTS~%
\lstinline[style=tlsf_default]~ {~\\%
\lstinline[style=tlsf_enumtyp]~    hburst~%
\lstinline[style=tlsf_variabl]~ HBURST~%
\lstinline[style=tlsf_default]~;~\\%
\lstinline[style=tlsf_default]~  }~\\%
\lstinline[style=tlsf_keyword]~  OUTPUTS~%
\lstinline[style=tlsf_default]~ {~\\%
\lstinline[style=tlsf_variabl]~    SINGLE~%
\lstinline[style=tlsf_default]~;~\\%
\lstinline[style=tlsf_variabl]~    BURST4~%
\lstinline[style=tlsf_default]~;~\\%
\lstinline[style=tlsf_variabl]~    INCR~%
\lstinline[style=tlsf_default]~;~\\%
\lstinline[style=tlsf_default]~    }~\\%
\lstinline[style=tlsf_keyword]~  ASSERT~%
\lstinline[style=tlsf_default]~ {~\\%
\lstinline[style=tlsf_default]~    HBURST ~%
\lstinline[style=tlsf_operato]~==~%
\lstinline[style=tlsf_enumval]~ Single ~%
\lstinline[style=tlsf_operato]~->~%
\lstinline[style=tlsf_default]~ SINGLE;~\\%
\lstinline[style=tlsf_default]~    HBURST ~%
\lstinline[style=tlsf_operato]~==~%
\lstinline[style=tlsf_enumval]~ Burst4 ~%
\lstinline[style=tlsf_operato]~->~%
\lstinline[style=tlsf_default]~ BURST4;~\\%
\lstinline[style=tlsf_default]~    HBURST ~%
\lstinline[style=tlsf_operato]~==~%
\lstinline[style=tlsf_enumval]~ Incr ~%
\lstinline[style=tlsf_operato]~->~%
\lstinline[style=tlsf_default]~ INCR;~\\%
\lstinline[style=tlsf_operato]~    !~%
\lstinline[style=tlsf_default]~(SINGLE ~%
\lstinline[style=tlsf_operato]~&&~%
\lstinline[style=tlsf_default]~ (BURST4 ~%
\lstinline[style=tlsf_operato]~||~%
\lstinline[style=tlsf_default]~ INCR)) ~%
\lstinline[style=tlsf_operato]~&&~%
\lstinline[style=tlsf_operato]~ !~%
\lstinline[style=tlsf_default]~(BURST4 ~%
\lstinline[style=tlsf_operato]~&&~%
\lstinline[style=tlsf_default]~ INCR);~\\%
\lstinline[style=tlsf_default]~  }~\\%
\lstinline[style=tlsf_default]~}~
\end{minipage}
\end{lrbox}

\newsavebox{\cArbiter}
\begin{lrbox}{\cArbiter}
\begin{minipage}{.49\textwidth}
\scriptsize
\lstinline[style=tlsf_keyword]~INFO~%
\lstinline[style=tlsf_default]~ {~\\%
\lstinline[style=tlsf_keyword]~  TITLE~%
\lstinline[style=tlsf_default]~:       ~%
\lstinline[style=tlsf_istring]~"AMBA AHB Arbiter"~\\%
\lstinline[style=tlsf_keyword]~  DESCRIPTION~%
\lstinline[style=tlsf_default]~: ~%
\lstinline[style=tlsf_istring]~"Component: Arbiter"~\\%
\lstinline[style=tlsf_keyword]~  SEMANTICS~%
\lstinline[style=tlsf_default]~:   ~%
\lstinline[style=tlsf_semanti]~Mealy~\\%
\lstinline[style=tlsf_keyword]~  TARGET~%
\lstinline[style=tlsf_default]~:      ~%
\lstinline[style=tlsf_semanti]~Mealy~\\%
\lstinline[style=tlsf_default]~}~\\%
\lstinline[style=tlsf_keyword]~GLOBAL ~%
\lstinline[style=tlsf_default]~{~\\%
\lstinline[style=tlsf_keyword]~  PARAMETERS ~%
\lstinline[style=tlsf_default]~{~\\%
\lstinline[style=tlsf_variabl]~    n~%
\lstinline[style=tlsf_default]~ = 2;~\\%
\lstinline[style=tlsf_default]~  }~\\%
\lstinline[style=tlsf_keyword]~  DEFINITIONS~%
\lstinline[style=tlsf_default]~ {~\\%
\lstinline[style=tlsf_comment]~    // mutual exclusion~\\%
\lstinline[style=tlsf_variabl]~    mutual~%
\lstinline[style=tlsf_default]~(b) =~\\%
\lstinline[style=tlsf_operato]~      ||~%
\lstinline[style=tlsf_default]~[i ~%
\lstinline[style=tlsf_operato]~IN~%
\lstinline[style=tlsf_default]~ {0, 1 .. (~%
\lstinline[style=tlsf_operato]~SIZEOF ~%
\lstinline[style=tlsf_default]~b)~%
\lstinline[style=tlsf_operato]~ - ~%
\lstinline[style=tlsf_default]~1}]~\\%
\lstinline[style=tlsf_operato]~        &&~%
\lstinline[style=tlsf_default]~[j ~%
\lstinline[style=tlsf_operato]~IN~%
\lstinline[style=tlsf_default]~ {0, 1 .. (~%
\lstinline[style=tlsf_operato]~SIZEOF ~%
\lstinline[style=tlsf_default]~b)~%
\lstinline[style=tlsf_operato]~ - ~%
\lstinline[style=tlsf_default]~1} ~%
\lstinline[style=tlsf_operato]~(\)~%
\lstinline[style=tlsf_default]~ {i}]~\\%
\lstinline[style=tlsf_operato]~          !~%
\lstinline[style=tlsf_default]~(b[i] ~%
\lstinline[style=tlsf_operato]~&&~%
\lstinline[style=tlsf_default]~ b[j]);~\\%
\lstinline[style=tlsf_default]~  }~\\%
\lstinline[style=tlsf_default]~}~\\%
\lstinline[style=tlsf_keyword]~MAIN~%
\lstinline[style=tlsf_default]~ {~\\%
\lstinline[style=tlsf_keyword]~  INPUTS~%
\lstinline[style=tlsf_default]~ {~\\%
\lstinline[style=tlsf_variabl]~    HBUSREQ~%
\lstinline[style=tlsf_default]~[n];~\\%
\lstinline[style=tlsf_variabl]~    ALLREADY~%
\lstinline[style=tlsf_default]~;~\\%
\lstinline[style=tlsf_default]~  }~\\%
\lstinline[style=tlsf_keyword]~  OUTPUTS~%
\lstinline[style=tlsf_default]~ {~\\%
\lstinline[style=tlsf_variabl]~    HGRANT~%
\lstinline[style=tlsf_default]~[n];~\\%
\lstinline[style=tlsf_variabl]~    BUSREQ~%
\lstinline[style=tlsf_default]~;~\\%
\lstinline[style=tlsf_variabl]~    DECIDE~%
\lstinline[style=tlsf_default]~;~\\%
\lstinline[style=tlsf_default]~  }~\\%
\lstinline[style=tlsf_keyword]~  INITIALLY~%
\lstinline[style=tlsf_default]~ {~\\%
\lstinline[style=tlsf_comment]~    // the component is initially idle~\\%
\lstinline[style=tlsf_default]~    ALLREADY;~\\%
\lstinline[style=tlsf_default]~  }~\\%
\lstinline[style=tlsf_keyword]~  ASSUME~%
\lstinline[style=tlsf_default]~ {~\\%
\lstinline[style=tlsf_comment]~    // the component is not eventually disabled~\\%
\lstinline[style=tlsf_operato]~    G F~%
\lstinline[style=tlsf_default]~ ALLREADY;~\\%
\lstinline[style=tlsf_default]~  }~\\%
\lstinline[style=tlsf_keyword]~  ASSERT~%
\lstinline[style=tlsf_default]~ {~\\%
\lstinline[style=tlsf_comment]~    // always exactely one master is granted~\\%
\lstinline[style=tlsf_default]~    mutual(HGRANT) ~%
\lstinline[style=tlsf_operato]~&& ||~%
\lstinline[style=tlsf_default]~[0 ~%
\lstinline[style=tlsf_operato]~<=~%
\lstinline[style=tlsf_default]~ i ~%
\lstinline[style=tlsf_operato]~<~%
\lstinline[style=tlsf_default]~ n] HGRANT[i];~\\%
\lstinline[style=tlsf_comment]~    // if not ready, the grants stay unchanged~\\%
\lstinline[style=tlsf_operato]~    &&~%
\lstinline[style=tlsf_default]~[0 ~%
\lstinline[style=tlsf_operato]~<=~%
\lstinline[style=tlsf_default]~ i ~%
\lstinline[style=tlsf_operato]~<~%
\lstinline[style=tlsf_default]~ n]~\\%
\lstinline[style=tlsf_default]~      (~%
\lstinline[style=tlsf_operato]~!~%
\lstinline[style=tlsf_default]~ALLREADY ~%
\lstinline[style=tlsf_operato]~->~%
\lstinline[style=tlsf_default]~ (~%
\lstinline[style=tlsf_operato]~X~%
\lstinline[style=tlsf_default]~ HGRANT[i] ~%
\lstinline[style=tlsf_operato]~<->~%
\lstinline[style=tlsf_default]~ HGRANT[i]));~\\%
\lstinline[style=tlsf_comment]~    // every request is eventually granted~\\%
\lstinline[style=tlsf_operato]~    &&~%
\lstinline[style=tlsf_default]~[0 ~%
\lstinline[style=tlsf_operato]~<=~%
\lstinline[style=tlsf_default]~ i ~%
\lstinline[style=tlsf_operato]~<~%
\lstinline[style=tlsf_default]~ n]~\\%
\lstinline[style=tlsf_default]~      (HBUSREQ[i] ~%
\lstinline[style=tlsf_operato]~-> F~%
\lstinline[style=tlsf_default]~ (~%
\lstinline[style=tlsf_operato]~!~%
\lstinline[style=tlsf_default]~HBUSREQ[i] ~%
\lstinline[style=tlsf_operato]~||~%
\lstinline[style=tlsf_default]~ HGRANT[i]));~\\%
\lstinline[style=tlsf_comment]~    // the BUSREQ signal mirrors the HBUSREQ[i]~\\%
\lstinline[style=tlsf_comment]~    // signal of the currently granted master i~\\%
\lstinline[style=tlsf_operato]~    &&~%
\lstinline[style=tlsf_default]~[0 ~%
\lstinline[style=tlsf_operato]~<=~%
\lstinline[style=tlsf_default]~ i ~%
\lstinline[style=tlsf_operato]~<~%
\lstinline[style=tlsf_default]~ n]~\\%
\lstinline[style=tlsf_default]~      (HGRANT[i] ~%
\lstinline[style=tlsf_operato]~->~%
\lstinline[style=tlsf_default]~ (BUSREQ ~%
\lstinline[style=tlsf_operato]~<->~%
\lstinline[style=tlsf_default]~ HBUSREQ[i]));~\\
\lstinline[style=tlsf_comment]~    // taking decisions requires to be idle~\\%
\lstinline[style=tlsf_operato]~    !~%
\lstinline[style=tlsf_default]~ALLREADY ~%
\lstinline[style=tlsf_operato]~-> !~%
\lstinline[style=tlsf_default]~DECIDE;~\\%
\lstinline[style=tlsf_comment]~    // granting another master triggers a decision~\\%
\lstinline[style=tlsf_default]~    DECIDE ~%
\lstinline[style=tlsf_operato]~<-> ||~%
\lstinline[style=tlsf_default]~[0 ~%
\lstinline[style=tlsf_operato]~<=~%
\lstinline[style=tlsf_default]~ i ~%
\lstinline[style=tlsf_operato]~<~%
\lstinline[style=tlsf_default]~ n]~\\%
\lstinline[style=tlsf_operato]~      !~%
\lstinline[style=tlsf_default]~(~%
\lstinline[style=tlsf_operato]~X~%
\lstinline[style=tlsf_default]~ HGRANT[i] ~%
\lstinline[style=tlsf_operato]~<->~%
\lstinline[style=tlsf_default]~ HGRANT[i]);~\\%
\lstinline[style=tlsf_comment]~    // if there is no request, master 0 is granted~\\%
\lstinline[style=tlsf_default]~    (~%
\lstinline[style=tlsf_operato]~&&~%
\lstinline[style=tlsf_default]~[0 ~%
\lstinline[style=tlsf_operato]~<=~%
\lstinline[style=tlsf_default]~ i ~%
\lstinline[style=tlsf_operato]~<~%
\lstinline[style=tlsf_default]~ n] ~%
\lstinline[style=tlsf_operato]~!~%
\lstinline[style=tlsf_default]~HBUSREQ[i]) ~%
\lstinline[style=tlsf_operato]~&&~%
\lstinline[style=tlsf_default]~ DECIDE~\\%
\lstinline[style=tlsf_operato]~       -> X~%
\lstinline[style=tlsf_default]~ HGRANT[0];~\\%
\lstinline[style=tlsf_default]~  }~\\%
\lstinline[style=tlsf_default]~}~
\end{minipage}
\end{lrbox}

\newsavebox{\cEncode}
\begin{lrbox}{\cEncode}
\begin{minipage}{.49\textwidth}
\scriptsize
\lstinline[style=tlsf_keyword]~INFO~%
\lstinline[style=tlsf_default]~ {~\\%
\lstinline[style=tlsf_keyword]~  TITLE~%
\lstinline[style=tlsf_default]~:       ~%
\lstinline[style=tlsf_istring]~"AMBA AHB Arbiter"~\\%
\lstinline[style=tlsf_keyword]~  DESCRIPTION~%
\lstinline[style=tlsf_default]~: ~%
\lstinline[style=tlsf_istring]~"Component: Encode"~\\%
\lstinline[style=tlsf_keyword]~  SEMANTICS~%
\lstinline[style=tlsf_default]~:   ~%
\lstinline[style=tlsf_semanti]~Mealy~\\%
\lstinline[style=tlsf_keyword]~  TARGET~%
\lstinline[style=tlsf_default]~:      ~%
\lstinline[style=tlsf_semanti]~Mealy~\\%
\lstinline[style=tlsf_default]~}~\\%
\lstinline[style=tlsf_keyword]~GLOBAL ~%
\lstinline[style=tlsf_default]~{~\\%
\lstinline[style=tlsf_keyword]~  PARAMETERS ~%
\lstinline[style=tlsf_default]~{~\\%
\lstinline[style=tlsf_variabl]~    n~%
\lstinline[style=tlsf_default]~ = 2;~\\%
\lstinline[style=tlsf_default]~  }~\\%
\lstinline[style=tlsf_keyword]~  DEFINITIONS~%
\lstinline[style=tlsf_default]~ {~\\%
\lstinline[style=tlsf_comment]~    // mutual exclusion~\\%
\lstinline[style=tlsf_variabl]~    mutual~%
\lstinline[style=tlsf_default]~(b) =~\\%
\lstinline[style=tlsf_operato]~      ||~%
\lstinline[style=tlsf_default]~[i ~%
\lstinline[style=tlsf_operato]~IN~%
\lstinline[style=tlsf_default]~ {0, 1 .. (~%
\lstinline[style=tlsf_operato]~SIZEOF ~%
\lstinline[style=tlsf_default]~b)~%
\lstinline[style=tlsf_operato]~ - ~%
\lstinline[style=tlsf_default]~1}]~\\%
\lstinline[style=tlsf_operato]~        &&~%
\lstinline[style=tlsf_default]~[j ~%
\lstinline[style=tlsf_operato]~IN~%
\lstinline[style=tlsf_default]~ {0, 1 .. (~%
\lstinline[style=tlsf_operato]~SIZEOF ~%
\lstinline[style=tlsf_default]~b)~%
\lstinline[style=tlsf_operato]~ - ~%
\lstinline[style=tlsf_default]~1} ~%
\lstinline[style=tlsf_operato]~(\)~%
\lstinline[style=tlsf_default]~ {i}]~\\%
\lstinline[style=tlsf_operato]~          !~%
\lstinline[style=tlsf_default]~(b[i] ~%
\lstinline[style=tlsf_operato]~&&~%
\lstinline[style=tlsf_default]~ b[j]);~\\%
\lstinline[style=tlsf_comment]~    // checks whether a bus encodes the numerical~\\%
\lstinline[style=tlsf_comment]~    // value v in binary~\\%
\lstinline[style=tlsf_variabl]~    value~%
\lstinline[style=tlsf_default]~(bus,v) = value'(bus,v,0,~
\lstinline[style=tlsf_operato]~SIZEOF~
\lstinline[style=tlsf_default]~ bus);~\\%
\lstinline[style=tlsf_variabl]~    value'~%
\lstinline[style=tlsf_default]~(bus,v,i,j) =~\\%
\lstinline[style=tlsf_default]~      i ~%
\lstinline[style=tlsf_operato]~>=~%
\lstinline[style=tlsf_default]~ j        : ~%
\lstinline[style=tlsf_keyword]~true~\\%
\lstinline[style=tlsf_default]~      bit(v,i) ~%
\lstinline[style=tlsf_operato]~==~%
\lstinline[style=tlsf_default]~ 1 : value'(bus,v,i~%
\lstinline[style=tlsf_operato]~+~%
\lstinline[style=tlsf_default]~1,j)~\\%
\lstinline[style=tlsf_operato]~                      &&~%
\lstinline[style=tlsf_default]~ bus[i]~\\%
\lstinline[style=tlsf_keyword]~      otherwise~%
\lstinline[style=tlsf_default]~     : value'(bus,v,i~%
\lstinline[style=tlsf_operato]~+~%
\lstinline[style=tlsf_default]~1,j)~\\%
\lstinline[style=tlsf_operato]~                      && !~%
\lstinline[style=tlsf_default]~bus[i];~\\%
\lstinline[style=tlsf_comment]~    // returns the i-th bit of the numerical~\\%
\lstinline[style=tlsf_comment]~    // value v~\\%
\lstinline[style=tlsf_variabl]~    bit~%
\lstinline[style=tlsf_default]~(v,i) =~\\%
\lstinline[style=tlsf_default]~      i ~%
\lstinline[style=tlsf_operato]~<=~%
\lstinline[style=tlsf_default]~ 0    : v ~%
\lstinline[style=tlsf_operato]~
\lstinline[style=tlsf_default]~ 2~\\%
\lstinline[style=tlsf_keyword]~      otherwise~%
\lstinline[style=tlsf_default]~ : bit(v~%
\lstinline[style=tlsf_operato]~/~%
\lstinline[style=tlsf_default]~2,i~%
\lstinline[style=tlsf_operato]~-~%
\lstinline[style=tlsf_default]~1);~\\%
\lstinline[style=tlsf_comment]~    // discrete logarithm~\\%
\lstinline[style=tlsf_variabl]~    log2~%
\lstinline[style=tlsf_default]~(x) =~\\%
\lstinline[style=tlsf_default]~      x ~%
\lstinline[style=tlsf_operato]~<=~%
\lstinline[style=tlsf_default]~ 1    : 1~\\%
\lstinline[style=tlsf_keyword]~      otherwise~%
\lstinline[style=tlsf_default]~ : 1 ~%
\lstinline[style=tlsf_operato]~+~%
\lstinline[style=tlsf_default]~ log2(x~%
\lstinline[style=tlsf_operato]~/~%
\lstinline[style=tlsf_default]~2);~\\%
\lstinline[style=tlsf_default]~  }~\\%
\lstinline[style=tlsf_default]~}~\\%
\lstinline[style=tlsf_keyword]~MAIN~%
\lstinline[style=tlsf_default]~ {~\\%
\lstinline[style=tlsf_keyword]~  INPUTS~%
\lstinline[style=tlsf_default]~ {~\\%
\lstinline[style=tlsf_variabl]~    HREADY~%
\lstinline[style=tlsf_default]~;~\\%
\lstinline[style=tlsf_variabl]~    HGRANT~%
\lstinline[style=tlsf_default]~[n];~\\%
\lstinline[style=tlsf_default]~  }~\\%
\lstinline[style=tlsf_keyword]~  OUTPUTS~%
\lstinline[style=tlsf_default]~ {~\\%
\lstinline[style=tlsf_comment]~    // the output is encoded in binary~\\%
\lstinline[style=tlsf_variabl]~    HMASTER~%
\lstinline[style=tlsf_default]~[log2(n~%
\lstinline[style=tlsf_operato]~-~%
\lstinline[style=tlsf_default]~1)];~\\%
\lstinline[style=tlsf_default]~  }~\\%
\lstinline[style=tlsf_keyword]~  REQUIRE~%
\lstinline[style=tlsf_default]~ {~\\%
\lstinline[style=tlsf_comment]~    // a every time exactely one grant is high~\\%
\lstinline[style=tlsf_default]~    mutual(HGRANT) ~%
\lstinline[style=tlsf_operato]~&& ||~%
\lstinline[style=tlsf_default]~[0 ~%
\lstinline[style=tlsf_operato]~<=~%
\lstinline[style=tlsf_default]~ i ~%
\lstinline[style=tlsf_operato]~<~%
\lstinline[style=tlsf_default]~ n] HGRANT[i];~\\%
\lstinline[style=tlsf_default]~  }~\\%
\lstinline[style=tlsf_keyword]~  ASSERT~%
\lstinline[style=tlsf_default]~ {~\\%
\lstinline[style=tlsf_comment]~    // output the binary encoding of i, whenever~\\%
\lstinline[style=tlsf_comment]~    // i is granted and HREADY is high~\\%
\lstinline[style=tlsf_operato]~    &&~%
\lstinline[style=tlsf_default]~[0 ~%
\lstinline[style=tlsf_operato]~<=~%
\lstinline[style=tlsf_default]~ i ~%
\lstinline[style=tlsf_operato]~<~%
\lstinline[style=tlsf_default]~ n] (HREADY ~%
\lstinline[style=tlsf_operato]~->~\\%
\lstinline[style=tlsf_default]~      (~%
\lstinline[style=tlsf_operato]~X~%
\lstinline[style=tlsf_default]~ value(HMASTER,i) ~%
\lstinline[style=tlsf_operato]~<->~%
\lstinline[style=tlsf_default]~ HGRANT[i]));~\\%
\lstinline[style=tlsf_comment]~    // when HREADY is low, the value is copied~\\%
\lstinline[style=tlsf_operato]~    !~%
\lstinline[style=tlsf_default]~HREADY ~%
\lstinline[style=tlsf_operato]~->~%
\lstinline[style=tlsf_operato]~ &&~%
\lstinline[style=tlsf_default]~[0 ~%
\lstinline[style=tlsf_operato]~<=~%
\lstinline[style=tlsf_default]~ i ~%
\lstinline[style=tlsf_operato]~<~%
\lstinline[style=tlsf_default]~ log2(n~%
\lstinline[style=tlsf_operato]~-~%
\lstinline[style=tlsf_default]~1)]~\\%
\lstinline[style=tlsf_default]~      (~%
\lstinline[style=tlsf_operato]~X~%
\lstinline[style=tlsf_default]~ HMASTER[i] ~%
\lstinline[style=tlsf_operato]~<->~%
\lstinline[style=tlsf_default]~ HMASTER[i]);~\\%
\lstinline[style=tlsf_default]~  }~\\%
\lstinline[style=tlsf_default]~}~
\end{minipage}
\end{lrbox}

\newsavebox{\cShift}
\begin{lrbox}{\cShift}
\begin{minipage}{0.9\textwidth}
\scriptsize
\lstinline[style=tlsf_keyword]~INFO~%
\lstinline[style=tlsf_default]~ {~\\%
\lstinline[style=tlsf_keyword]~  TITLE~%
\lstinline[style=tlsf_default]~:       ~%
\lstinline[style=tlsf_istring]~"AMBA AHB Arbiter"~\\%
\lstinline[style=tlsf_keyword]~  DESCRIPTION~%
\lstinline[style=tlsf_default]~: ~%
\lstinline[style=tlsf_istring]~"Component: Shift"~\\%
\lstinline[style=tlsf_keyword]~  SEMANTICS~%
\lstinline[style=tlsf_default]~:   ~%
\lstinline[style=tlsf_semanti]~Mealy~\\%
\lstinline[style=tlsf_keyword]~  TARGET~%
\lstinline[style=tlsf_default]~:      ~%
\lstinline[style=tlsf_semanti]~Mealy~\\%
\lstinline[style=tlsf_default]~}~\\%
\lstinline[style=tlsf_keyword]~MAIN~%
\lstinline[style=tlsf_default]~ {~\\%
\lstinline[style=tlsf_keyword]~  INPUTS~%
\lstinline[style=tlsf_default]~ { ~%
\lstinline[style=tlsf_variabl]~HREADY~%
\lstinline[style=tlsf_default]~; ~%
\lstinline[style=tlsf_variabl]~LOCKED~%
\lstinline[style=tlsf_default]~;~%
\lstinline[style=tlsf_default]~ }~\\%
\lstinline[style=tlsf_keyword]~  OUTPUTS~%
\lstinline[style=tlsf_default]~ { ~%
\lstinline[style=tlsf_variabl]~HMASTLOCK~%
\lstinline[style=tlsf_default]~;~%
\lstinline[style=tlsf_default]~ }~\\%
\lstinline[style=tlsf_keyword]~  ASSERT~%
\lstinline[style=tlsf_default]~ {~\\%
\lstinline[style=tlsf_comment]~    // if HREADY is high, the component copies LOCKED to HMASTLOCK, shifted by one time step~\\%
\lstinline[style=tlsf_default]~    HREADY ~%
\lstinline[style=tlsf_operato]~->~%
\lstinline[style=tlsf_default]~ (~%
\lstinline[style=tlsf_operato]~X~%
\lstinline[style=tlsf_default]~ HMASTLOCK ~%
\lstinline[style=tlsf_operato]~<->~%
\lstinline[style=tlsf_default]~ LOCKED);~\\%
\lstinline[style=tlsf_comment]~    // if HREADY is low, the old value of HMASTLOCK is copied~\\%
\lstinline[style=tlsf_operato]~    !~%
\lstinline[style=tlsf_default]~HREADY ~%
\lstinline[style=tlsf_operato]~->~%
\lstinline[style=tlsf_default]~ (~%
\lstinline[style=tlsf_operato]~X~%
\lstinline[style=tlsf_default]~ HMASTLOCK ~%
\lstinline[style=tlsf_operato]~<->~%
\lstinline[style=tlsf_default]~ HMASTLOCK);~\\%
\lstinline[style=tlsf_default]~  }~\\%
\lstinline[style=tlsf_default]~}~
\end{minipage}
\end{lrbox}

\newsavebox{\cTincr}
\begin{lrbox}{\cTincr}
\begin{minipage}{0.9\textwidth}
\scriptsize
\lstinline[style=tlsf_keyword]~INFO~%
\lstinline[style=tlsf_default]~ {~\\%
\lstinline[style=tlsf_keyword]~  TITLE~%
\lstinline[style=tlsf_default]~:       ~%
\lstinline[style=tlsf_istring]~"AMBA AHB Arbiter"~\\%
\lstinline[style=tlsf_keyword]~  DESCRIPTION~%
\lstinline[style=tlsf_default]~: ~%
\lstinline[style=tlsf_istring]~"Component: TIncr"~\\%
\lstinline[style=tlsf_keyword]~  SEMANTICS~%
\lstinline[style=tlsf_default]~:   ~%
\lstinline[style=tlsf_semanti]~Mealy~\\%
\lstinline[style=tlsf_keyword]~  TARGET~%
\lstinline[style=tlsf_default]~:      ~%
\lstinline[style=tlsf_semanti]~Mealy~\\%
\lstinline[style=tlsf_default]~}~\\%
\lstinline[style=tlsf_keyword]~MAIN~%
\lstinline[style=tlsf_default]~ {~\\%
\lstinline[style=tlsf_keyword]~  INPUTS~%
\lstinline[style=tlsf_default]~ { ~%
\lstinline[style=tlsf_variabl]~INCR~%
\lstinline[style=tlsf_default]~; ~%
\lstinline[style=tlsf_variabl]~HREADY~%
\lstinline[style=tlsf_default]~; ~%
\lstinline[style=tlsf_variabl]~LOCKED~%
\lstinline[style=tlsf_default]~; ~%
\lstinline[style=tlsf_variabl]~DECIDE~%
\lstinline[style=tlsf_default]~; ~%
\lstinline[style=tlsf_variabl]~BUSREQ~%
\lstinline[style=tlsf_default]~; ~%
\lstinline[style=tlsf_default]~}~\\%
\lstinline[style=tlsf_keyword]~  OUTPUTS~%
\lstinline[style=tlsf_default]~ { ~%
\lstinline[style=tlsf_variabl]~READY1~%
\lstinline[style=tlsf_default]~;~%
\lstinline[style=tlsf_default]~ }~\\%
\lstinline[style=tlsf_keyword]~  INITIALLY~%
\lstinline[style=tlsf_default]~ { ~%
\lstinline[style=tlsf_operato]~!~%
\lstinline[style=tlsf_default]~DECIDE;~%
\lstinline[style=tlsf_default]~ }~\\%
\lstinline[style=tlsf_keyword]~  PRESET~%
\lstinline[style=tlsf_default]~ { ~%
\lstinline[style=tlsf_default]~READY1;~%
\lstinline[style=tlsf_default]~ }~\\%
\lstinline[style=tlsf_keyword]~  REQUIRE~%
\lstinline[style=tlsf_default]~ {~\\%
\lstinline[style=tlsf_comment]~    // decisions are only taken if the component is ready~\\%
\lstinline[style=tlsf_operato]~    !~%
\lstinline[style=tlsf_default]~READY1 ~%
\lstinline[style=tlsf_operato]~-> X !~%
\lstinline[style=tlsf_default]~DECIDE;~\\%
\lstinline[style=tlsf_default]~  }~\\%
\lstinline[style=tlsf_keyword]~  ASSUME~%
\lstinline[style=tlsf_default]~ {~\\%
\lstinline[style=tlsf_comment]~    // slaves and masters cannot block the bus~\\%
\lstinline[style=tlsf_operato]~    G F ~%
\lstinline[style=tlsf_default]~HREADY ~%
\lstinline[style=tlsf_operato]~&& G F !~%
\lstinline[style=tlsf_default]~BUSREQ;~\\%
\lstinline[style=tlsf_default]~  }~\\%
\lstinline[style=tlsf_keyword]~  ASSERT~%
\lstinline[style=tlsf_default]~ {~\\%
\lstinline[style=tlsf_comment]~    // for each incremental, locked transmission, the bus is locked as long as requested~\\%
\lstinline[style=tlsf_default]~    DECIDE ~%
\lstinline[style=tlsf_operato]~->~\\%
\lstinline[style=tlsf_operato]~      X~%
\lstinline[style=tlsf_default]~[2] (((INCR ~%
\lstinline[style=tlsf_operato]~&&~%
\lstinline[style=tlsf_default]~ LOCKED) ~%
\lstinline[style=tlsf_operato]~->~%
\lstinline[style=tlsf_default]~ (~%
\lstinline[style=tlsf_operato]~!~%
\lstinline[style=tlsf_default]~READY1 ~%
\lstinline[style=tlsf_operato]~W~%
\lstinline[style=tlsf_default]~ (HREADY ~%
\lstinline[style=tlsf_operato]~&& !~
\lstinline[style=tlsf_default]~BUSREQ))) ~%
\lstinline[style=tlsf_operato]~&&~\\%
\lstinline[style=tlsf_default]~            (~%
\lstinline[style=tlsf_operato]~!~%
\lstinline[style=tlsf_default]~(INCR ~%
\lstinline[style=tlsf_operato]~&&~%
\lstinline[style=tlsf_default]~ LOCKED) ~%
\lstinline[style=tlsf_operato]~->~%
\lstinline[style=tlsf_default]~ READY1));~\\%
\lstinline[style=tlsf_comment]~    // the component stays ready as long as there is no decision~\\%
\lstinline[style=tlsf_default]~    READY ~%
\lstinline[style=tlsf_operato]~&& X !~%
\lstinline[style=tlsf_default]~DECIDE ~%
\lstinline[style=tlsf_operato]~-> X~%
\lstinline[style=tlsf_default]~ READY1;~\\%
\lstinline[style=tlsf_comment]~    // if there is a decision the component blocks the bus for at least two time steps~\\%
\lstinline[style=tlsf_default]~    READY1 ~%
\lstinline[style=tlsf_operato]~&& X~%
\lstinline[style=tlsf_default]~ DECIDE ~%
\lstinline[style=tlsf_operato]~-> G~%
\lstinline[style=tlsf_default]~[1:2] ~%
\lstinline[style=tlsf_operato]~!~%
\lstinline[style=tlsf_default]~ READY1;~\\%
\lstinline[style=tlsf_default]~  }~\\%
\lstinline[style=tlsf_default]~}~
\end{minipage}
\end{lrbox}

\newsavebox{\cSingle}
\begin{lrbox}{\cSingle}
\begin{minipage}{0.9\textwidth}
\scriptsize
\lstinline[style=tlsf_keyword]~INFO~%
\lstinline[style=tlsf_default]~ {~\\%
\lstinline[style=tlsf_keyword]~  TITLE~%
\lstinline[style=tlsf_default]~:       ~%
\lstinline[style=tlsf_istring]~"AMBA AHB Arbiter"~\\%
\lstinline[style=tlsf_keyword]~  DESCRIPTION~%
\lstinline[style=tlsf_default]~: ~%
\lstinline[style=tlsf_istring]~"Component: TSingle"~\\%
\lstinline[style=tlsf_keyword]~  SEMANTICS~%
\lstinline[style=tlsf_default]~:   ~%
\lstinline[style=tlsf_semanti]~Mealy~\\%
\lstinline[style=tlsf_keyword]~  TARGET~%
\lstinline[style=tlsf_default]~:      ~%
\lstinline[style=tlsf_semanti]~Mealy~\\%
\lstinline[style=tlsf_default]~}~\\%
\lstinline[style=tlsf_keyword]~MAIN~%
\lstinline[style=tlsf_default]~ {~\\%
\lstinline[style=tlsf_keyword]~  INPUTS~%
\lstinline[style=tlsf_default]~ { ~%
\lstinline[style=tlsf_variabl]~SINGLE~%
\lstinline[style=tlsf_default]~; ~%
\lstinline[style=tlsf_variabl]~HREADY~%
\lstinline[style=tlsf_default]~; ~%
\lstinline[style=tlsf_variabl]~LOCKED~%
\lstinline[style=tlsf_default]~; ~%
\lstinline[style=tlsf_variabl]~DECIDE~%
\lstinline[style=tlsf_default]~; ~%
\lstinline[style=tlsf_default]~}~\\%
\lstinline[style=tlsf_keyword]~  OUTPUTS~%
\lstinline[style=tlsf_default]~ { ~%
\lstinline[style=tlsf_variabl]~READY3~%
\lstinline[style=tlsf_default]~;~%
\lstinline[style=tlsf_default]~ }~\\%
\lstinline[style=tlsf_keyword]~  INITIALLY~%
\lstinline[style=tlsf_default]~ {~\\%
\lstinline[style=tlsf_comment]~    // initially no decision is taken~\\%
\lstinline[style=tlsf_operato]~    !~%
\lstinline[style=tlsf_default]~DECIDE;~\\%
\lstinline[style=tlsf_default]~  }~\\%
\lstinline[style=tlsf_keyword]~  PRESET~%
\lstinline[style=tlsf_default]~ {~\\%
\lstinline[style=tlsf_comment]~    // at startup, the component is ready~\\%
\lstinline[style=tlsf_default]~    READY3;~\\%
\lstinline[style=tlsf_default]~  }~\\%
\lstinline[style=tlsf_keyword]~  REQUIRE~%
\lstinline[style=tlsf_default]~ {~\\%
\lstinline[style=tlsf_comment]~    // decisions are only taken if the component is ready~\\%
\lstinline[style=tlsf_operato]~    !~%
\lstinline[style=tlsf_default]~READY3 ~%
\lstinline[style=tlsf_operato]~-> X !~%
\lstinline[style=tlsf_default]~DECIDE;~\\%
\lstinline[style=tlsf_default]~  }~\\%
\lstinline[style=tlsf_keyword]~  ASSUME~%
\lstinline[style=tlsf_default]~ {~\\%
\lstinline[style=tlsf_comment]~    // a slave cannot block the bus~\\%
\lstinline[style=tlsf_operato]~    G F ~%
\lstinline[style=tlsf_default]~HREADY ~\\%
\lstinline[style=tlsf_default]~  }~\\%
\lstinline[style=tlsf_keyword]~  ASSERT~%
\lstinline[style=tlsf_default]~ {~\\%
\lstinline[style=tlsf_comment]~    // for each single, locked transmission, the bus is locked for one time step~\\%
\lstinline[style=tlsf_default]~    DECIDE ~%
\lstinline[style=tlsf_operato]~->~\\%
\lstinline[style=tlsf_operato]~      X~%
\lstinline[style=tlsf_default]~[2] (((SINGLE ~%
\lstinline[style=tlsf_operato]~&&~%
\lstinline[style=tlsf_default]~ LOCKED) ~%
\lstinline[style=tlsf_operato]~->~%
\lstinline[style=tlsf_default]~ (~%
\lstinline[style=tlsf_operato]~!~%
\lstinline[style=tlsf_default]~READY3 ~%
\lstinline[style=tlsf_operato]~U~%
\lstinline[style=tlsf_default]~ (HREADY ~%
\lstinline[style=tlsf_operato]~&& !~
\lstinline[style=tlsf_default]~READY3 ~%
\lstinline[style=tlsf_operato]~&& X~%
\lstinline[style=tlsf_default]~ READY3))) ~%
\lstinline[style=tlsf_operato]~&&~\\%
\lstinline[style=tlsf_default]~            (~%
\lstinline[style=tlsf_operato]~!~%
\lstinline[style=tlsf_default]~(SINGLE ~%
\lstinline[style=tlsf_operato]~&&~%
\lstinline[style=tlsf_default]~ LOCKED) ~%
\lstinline[style=tlsf_operato]~->~%
\lstinline[style=tlsf_default]~ READY3));~\\%
\lstinline[style=tlsf_comment]~    // the component stays ready as long as there is no decision~\\%
\lstinline[style=tlsf_default]~    READY3 ~%
\lstinline[style=tlsf_operato]~&& X !~%
\lstinline[style=tlsf_default]~DECIDE ~%
\lstinline[style=tlsf_operato]~-> X~%
\lstinline[style=tlsf_default]~ READY3;~\\%
\lstinline[style=tlsf_comment]~    // if there is a decision the component blocks the bus for at least two time steps~\\%
\lstinline[style=tlsf_default]~    READY3 ~%
\lstinline[style=tlsf_operato]~&& X~%
\lstinline[style=tlsf_default]~ DECIDE ~%
\lstinline[style=tlsf_operato]~-> G~%
\lstinline[style=tlsf_default]~[1:2] ~%
\lstinline[style=tlsf_operato]~!~%
\lstinline[style=tlsf_default]~ READY3;~\\%
\lstinline[style=tlsf_default]~  }~\\%
\lstinline[style=tlsf_default]~}~
\end{minipage}
\end{lrbox}

\newsavebox{\cBurst}
\begin{lrbox}{\cBurst}
\begin{minipage}{0.9\textwidth}
\scriptsize
\lstinline[style=tlsf_keyword]~INFO~%
\lstinline[style=tlsf_default]~ {~\\%
\lstinline[style=tlsf_keyword]~  TITLE~%
\lstinline[style=tlsf_default]~:       ~%
\lstinline[style=tlsf_istring]~"AMBA AHB Arbiter"~\\%
\lstinline[style=tlsf_keyword]~  DESCRIPTION~%
\lstinline[style=tlsf_default]~: ~%
\lstinline[style=tlsf_istring]~"Component: TBurst4"~\\%
\lstinline[style=tlsf_keyword]~  SEMANTICS~%
\lstinline[style=tlsf_default]~:   ~%
\lstinline[style=tlsf_semanti]~Mealy~\\%
\lstinline[style=tlsf_keyword]~  TARGET~%
\lstinline[style=tlsf_default]~:      ~%
\lstinline[style=tlsf_semanti]~Mealy~\\%
\lstinline[style=tlsf_default]~}~\\%
\lstinline[style=tlsf_keyword]~MAIN~%
\lstinline[style=tlsf_default]~ {~\\%
\lstinline[style=tlsf_keyword]~  INPUTS~%
\lstinline[style=tlsf_default]~ { ~%
\lstinline[style=tlsf_variabl]~BURST4~%
\lstinline[style=tlsf_default]~; ~%
\lstinline[style=tlsf_variabl]~HREADY~%
\lstinline[style=tlsf_default]~; ~%
\lstinline[style=tlsf_variabl]~LOCKED~%
\lstinline[style=tlsf_default]~; ~%
\lstinline[style=tlsf_variabl]~DECIDE~%
\lstinline[style=tlsf_default]~; ~%
\lstinline[style=tlsf_default]~}~\\%
\lstinline[style=tlsf_keyword]~  OUTPUTS~%
\lstinline[style=tlsf_default]~ { ~%
\lstinline[style=tlsf_variabl]~READY2~%
\lstinline[style=tlsf_default]~;~%
\lstinline[style=tlsf_default]~ }~\\%
\lstinline[style=tlsf_keyword]~  INITIALLY~%
\lstinline[style=tlsf_default]~ { ~%
\lstinline[style=tlsf_operato]~!~%
\lstinline[style=tlsf_default]~DECIDE;~%
\lstinline[style=tlsf_default]~ }~\\%
\lstinline[style=tlsf_keyword]~  PRESET~%
\lstinline[style=tlsf_default]~ { ~%
\lstinline[style=tlsf_default]~READY2;~%
\lstinline[style=tlsf_default]~ }~\\%
\lstinline[style=tlsf_keyword]~  REQUIRE~%
\lstinline[style=tlsf_default]~ {~\\%
\lstinline[style=tlsf_comment]~    // decisions are only taken if the component is ready~\\%
\lstinline[style=tlsf_operato]~    !~%
\lstinline[style=tlsf_default]~READY2 ~%
\lstinline[style=tlsf_operato]~-> X !~%
\lstinline[style=tlsf_default]~DECIDE;~\\%
\lstinline[style=tlsf_default]~  }~\\%
\lstinline[style=tlsf_keyword]~  ASSUME~%
\lstinline[style=tlsf_default]~ {~\\%
\lstinline[style=tlsf_comment]~    // a slave block the bus~\\%
\lstinline[style=tlsf_operato]~    G F ~%
\lstinline[style=tlsf_default]~HREADY;~\\%
\lstinline[style=tlsf_default]~  }~\\%
\lstinline[style=tlsf_keyword]~  ASSERT~%
\lstinline[style=tlsf_default]~ {~\\%
\lstinline[style=tlsf_comment]~    // for each burst4, locked transmission, the bus is locked for four time steps~\\%
\lstinline[style=tlsf_default]~    DECIDE ~%
\lstinline[style=tlsf_operato]~->~\\%
\lstinline[style=tlsf_operato]~      X~%
\lstinline[style=tlsf_default]~[2] (((BURST4 ~%
\lstinline[style=tlsf_operato]~&&~%
\lstinline[style=tlsf_default]~ LOCKED) ~%
\lstinline[style=tlsf_operato]~->~%
\lstinline[style=tlsf_default]~ (~%
\lstinline[style=tlsf_operato]~!~%
\lstinline[style=tlsf_default]~READY2 ~%
\lstinline[style=tlsf_operato]~U~%
\lstinline[style=tlsf_default]~ (HREADY ~%
\lstinline[style=tlsf_operato]~&& !~
\lstinline[style=tlsf_default]~READY2 ~%
\lstinline[style=tlsf_operato]~&& X~%
\lstinline[style=tlsf_default]~ (~%
\lstinline[style=tlsf_operato]~!~%
\lstinline[style=tlsf_default]~READY2 ~%
\lstinline[style=tlsf_operato]~U~%
\lstinline[style=tlsf_default]~ (HREADY ~%
\lstinline[style=tlsf_operato]~&&~\\%
\lstinline[style=tlsf_operato]~             !~%
\lstinline[style=tlsf_default]~READY2 ~%
\lstinline[style=tlsf_operato]~&& X~%
\lstinline[style=tlsf_default]~ (~%
\lstinline[style=tlsf_operato]~!~%
\lstinline[style=tlsf_default]~READY2 ~%
\lstinline[style=tlsf_operato]~U~%
\lstinline[style=tlsf_default]~ (HREADY ~%
\lstinline[style=tlsf_operato]~&& !~%
\lstinline[style=tlsf_default]~READY2 ~%
\lstinline[style=tlsf_operato]~&& X~%
\lstinline[style=tlsf_default]~ (~%
\lstinline[style=tlsf_operato]~!~%
\lstinline[style=tlsf_default]~READY2 ~%
\lstinline[style=tlsf_operato]~U~%
\lstinline[style=tlsf_default]~ (HREADY ~%
\lstinline[style=tlsf_operato]~&&~\\%
\lstinline[style=tlsf_operato]~             !~%
\lstinline[style=tlsf_default]~READY2 ~%
\lstinline[style=tlsf_operato]~&& X~%
\lstinline[style=tlsf_default]~READY2))))))))) ~%
\lstinline[style=tlsf_operato]~&&~%
\lstinline[style=tlsf_default]~ (~%
\lstinline[style=tlsf_operato]~!~%
\lstinline[style=tlsf_default]~(BURST4 ~%
\lstinline[style=tlsf_operato]~&&~%
\lstinline[style=tlsf_default]~ LOCKED) ~%
\lstinline[style=tlsf_operato]~->~%
\lstinline[style=tlsf_default]~ READY2)) ~\\%
\lstinline[style=tlsf_comment]~    // the component stays ready as long as there is no decision~\\%
\lstinline[style=tlsf_default]~    READY2 ~%
\lstinline[style=tlsf_operato]~&& X !~%
\lstinline[style=tlsf_default]~DECIDE ~%
\lstinline[style=tlsf_operato]~-> X~%
\lstinline[style=tlsf_default]~ READY2;~\\%
\lstinline[style=tlsf_comment]~    // if there is a decision the component blocks the bus for at least two time steps~\\%
\lstinline[style=tlsf_default]~    READY2 ~%
\lstinline[style=tlsf_operato]~&& X~%
\lstinline[style=tlsf_default]~ DECIDE ~%
\lstinline[style=tlsf_operato]~-> G~%
\lstinline[style=tlsf_default]~[1:2] ~%
\lstinline[style=tlsf_operato]~!~%
\lstinline[style=tlsf_default]~ READY2;~\\%
\lstinline[style=tlsf_default]~  }~\\%
\lstinline[style=tlsf_default]~}~
\end{minipage}
\end{lrbox}

\newsavebox{\cLock}
\begin{lrbox}{\cLock}
\begin{minipage}{0.9\textwidth}
\scriptsize
\lstinline[style=tlsf_keyword]~INFO~%
\lstinline[style=tlsf_default]~ {~\\%
\lstinline[style=tlsf_keyword]~  TITLE~%
\lstinline[style=tlsf_default]~:       ~%
\lstinline[style=tlsf_istring]~"AMBA AHB Arbiter"~\\%
\lstinline[style=tlsf_keyword]~  DESCRIPTION~%
\lstinline[style=tlsf_default]~: ~%
\lstinline[style=tlsf_istring]~"Component: Lock"~\\%
\lstinline[style=tlsf_keyword]~  SEMANTICS~%
\lstinline[style=tlsf_default]~:   ~%
\lstinline[style=tlsf_semanti]~Mealy~\\%
\lstinline[style=tlsf_keyword]~  TARGET~%
\lstinline[style=tlsf_default]~:      ~%
\lstinline[style=tlsf_semanti]~Mealy~\\%
\lstinline[style=tlsf_default]~}~\\%
\lstinline[style=tlsf_keyword]~GLOBAL ~%
\lstinline[style=tlsf_default]~{~\\%
\lstinline[style=tlsf_keyword]~  PARAMETERS ~%
\lstinline[style=tlsf_default]~{~\\%
\lstinline[style=tlsf_variabl]~    n~%
\lstinline[style=tlsf_default]~ = 2;~\\%
\lstinline[style=tlsf_default]~  }~\\%
\lstinline[style=tlsf_keyword]~  DEFINITIONS~%
\lstinline[style=tlsf_default]~ {~\\%
\lstinline[style=tlsf_comment]~    // mutual exclusion~\\%
\lstinline[style=tlsf_variabl]~    mutual~%
\lstinline[style=tlsf_default]~(b) =~\\%
\lstinline[style=tlsf_operato]~      ||~%
\lstinline[style=tlsf_default]~[i ~%
\lstinline[style=tlsf_operato]~IN~%
\lstinline[style=tlsf_default]~ {0, 1 .. (~%
\lstinline[style=tlsf_operato]~SIZEOF ~%
\lstinline[style=tlsf_default]~b)~%
\lstinline[style=tlsf_operato]~ - ~%
\lstinline[style=tlsf_default]~1}]~\\%
\lstinline[style=tlsf_operato]~        &&~%
\lstinline[style=tlsf_default]~[j ~%
\lstinline[style=tlsf_operato]~IN~%
\lstinline[style=tlsf_default]~ {0, 1 .. (~%
\lstinline[style=tlsf_operato]~SIZEOF ~%
\lstinline[style=tlsf_default]~b)~%
\lstinline[style=tlsf_operato]~ - ~%
\lstinline[style=tlsf_default]~1} ~%
\lstinline[style=tlsf_operato]~(\)~%
\lstinline[style=tlsf_default]~ {i}]~\\%
\lstinline[style=tlsf_operato]~          !~%
\lstinline[style=tlsf_default]~(b[i] ~%
\lstinline[style=tlsf_operato]~&&~%
\lstinline[style=tlsf_default]~ b[j]);~\\%
\lstinline[style=tlsf_default]~  }~\\%
\lstinline[style=tlsf_default]~}~\\%
\lstinline[style=tlsf_keyword]~MAIN~%
\lstinline[style=tlsf_default]~ {~\\%
\lstinline[style=tlsf_keyword]~  INPUTS~%
\lstinline[style=tlsf_default]~ {~\\%
\lstinline[style=tlsf_variabl]~    DECIDE~%
\lstinline[style=tlsf_default]~;~\\%
\lstinline[style=tlsf_variabl]~    HGRANT~%
\lstinline[style=tlsf_default]~[n];~\\%
\lstinline[style=tlsf_variabl]~    HLOCK~%
\lstinline[style=tlsf_default]~[n];~\\%
\lstinline[style=tlsf_default]~  }~\\%
\lstinline[style=tlsf_keyword]~  OUTPUTS~%
\lstinline[style=tlsf_default]~ {~\\%
\lstinline[style=tlsf_variabl]~    LOCKED~%
\lstinline[style=tlsf_default]~;~\\%
\lstinline[style=tlsf_default]~  }~\\%
\lstinline[style=tlsf_keyword]~  REQUIRE~%
\lstinline[style=tlsf_default]~ {~\\%
\lstinline[style=tlsf_comment]~    // a every time exactely one grant is high~\\%
\lstinline[style=tlsf_default]~    mutual(HGRANT) ~%
\lstinline[style=tlsf_operato]~&& ||~%
\lstinline[style=tlsf_default]~[0 ~%
\lstinline[style=tlsf_operato]~<=~%
\lstinline[style=tlsf_default]~ i ~%
\lstinline[style=tlsf_operato]~<~%
\lstinline[style=tlsf_default]~ n] HGRANT[i];~\\%
\lstinline[style=tlsf_default]~  }~\\%
\lstinline[style=tlsf_keyword]~  ASSERT~%
\lstinline[style=tlsf_default]~ {~\\%
\lstinline[style=tlsf_comment]~    // whenever a decicion is taken, the LOCKED signal is updated to~\\%
\lstinline[style=tlsf_comment]~    // the HLOCK value of the granted master~\\%
\lstinline[style=tlsf_operato]~    &&~%
\lstinline[style=tlsf_default]~[0 ~%
\lstinline[style=tlsf_operato]~<=~%
\lstinline[style=tlsf_default]~ i ~%
\lstinline[style=tlsf_operato]~<~%
\lstinline[style=tlsf_default]~ n] (DECIDE ~%
\lstinline[style=tlsf_operato]~&& X~%
\lstinline[style=tlsf_default]~ HGRANT[i] ~%
\lstinline[style=tlsf_operato]~->~%
\lstinline[style=tlsf_default]~ (~%
\lstinline[style=tlsf_operato]~X~%
\lstinline[style=tlsf_default]~ LOCKED ~%
\lstinline[style=tlsf_operato]~<-> X~%
\lstinline[style=tlsf_default]~ HLOCK[i]));~\\%
\lstinline[style=tlsf_comment]~    // otherwise, the value is copied~\\%
\lstinline[style=tlsf_operato]~    !~%
\lstinline[style=tlsf_default]~DECIDE ~%
\lstinline[style=tlsf_operato]~->~%
\lstinline[style=tlsf_default]~ (~%
\lstinline[style=tlsf_operato]~X~%
\lstinline[style=tlsf_default]~ LOCKED ~%
\lstinline[style=tlsf_operato]~<->~%
\lstinline[style=tlsf_default]~ LOCKED);~\\%
\lstinline[style=tlsf_default]~  }~\\%
\lstinline[style=tlsf_default]~}~
\end{minipage}
\end{lrbox}

\maketitle

\begin{abstract}
  We present an extension of the Temporal Logic Synthesis Format (TLSF). 
TLSF builds on standard LTL, but additionally supports 
high-level constructs, such as sets and functions, as well as parameters 
that allow a specification to define a whole a family of problems. 
Our extension introduces operators and a new semantics option for \ltlf, 
i.e., LTL on finite executions.

\end{abstract}


We assume the reader is already familiar with the TLSF (v1.1) synthesis
format, please refer to~\cite{tlsfv1.1}.

\section{The Basic Format}
\label{sec:basicformat}
We first recap the basic format of TLSF v1.1 in sections \cref{sec:basicinfo,sec:basicmain,sec:basicexpr}, 
and then introduce extensions for \ltlf in \cref{sec:basicLTLf}.

A specification in the basic format consists of an \lstinline!INFO! section and a
\lstinline!MAIN! section:
\begin{equation*}
  \tlsfsec{info} \tlsfsec{main} 
\end{equation*}

\subsection{The INFO Section}
\label{sec:basicinfo}

The \lstinline!INFO! section contains the meta data of the specification, like a
title and some description\footnote{We use colored verbatim font to
  identify the syntactic elements of the specification.}. Furthermore,
it defines the underlying semantics of the specification (Mealy or
Moore / standard, strict implication, or finite semantics) and the target model of the
synthesized implementation. Detailed information about supported
semantics and targets can be found in \secref{sec:semantics}.
Finally, a comma separated list of tags can be specified to identify
features of the specification, e.g., the restriction to a specific
fragment of LTL. A \tlsfid{tag} can be any string literal and is not
restricted to any predefined keywords.

\goodbreak

\vspace{1em}

\noindent
\lstinline!  INFO {!\\%
\lstinline!    TITLE:       "!$ \tlsfid{some title} $\lstinline!"!\\%
\lstinline!    DESCRIPTION: "!$ \tlsfid{some description} $\lstinline!"!\\%
\lstinline!    SEMANTICS:   !$ \tlsfid{semantics} $\\%
\lstinline!    TARGET:      !$ \tlsfid{target} $\\%
\lstinline!    TAGS:        !$ \tlsfid{tag} $%
\lstinline!,! $ \tlsfid{tag} $\lstinline!,!$ \ \ldots $\\%
\lstinline!  }!

\subsection{The MAIN Section}
\label{sec:basicmain}

The specification is completed by the \lstinline!MAIN! section, which 
contains the
partitioning of input and output signals, followed by the main
specification. The specification itself is separated into assumptions on the 
environment and desired properties of the system, and can additionally be 
distinguished into initial (\lstinline!INITIALLY!/\lstinline!PRESET!), 
invariant (\lstinline!REQUIRE!/\lstinline!ASSERT!), and arbitrary 
(\lstinline!ASSUME!/\lstinline!GUARANTEE!) properties\footnote{In TLSF 
v1.0~\cite{JacobsK16}, \lstinline!ASSERT! was called \lstinline!INVARIANTS!, \lstinline!ASSUME! was called 
\lstinline!ASSUMPTIONS!, and \lstinline!GUARANTEE! was called \lstinline!GUARANTEES! (and subsections \lstinline!INITIALLY!, 
\lstinline!PRESET!, and \lstinline!REQUIRE! did not exist). TLSF v1.1 still supports the old 
identifiers.}.  Multiple declarations and expressions need
to be separated by a '\lstinline!;!'.

\vspace{1em}

\noindent
\lstinline!  MAIN {!\\%
\lstinline!    INPUTS    { !%
$ ( \tlsfsec{boolean signal declaration} $\lstinline!;!$ )^{*} $\lstinline! }!\\%
\lstinline!    OUTPUTS   { !%
$ ( \tlsfsec{boolean signal declaration} $\lstinline!;!$ )^{*} $\lstinline! }!\\%
\lstinline!    INITIALLY { !%
$ ( \tlsfsec{basic LTL expression} $\lstinline!;!$ )^{*} $\lstinline! }!\\%
\lstinline!    PRESET    { !%
$ ( \tlsfsec{basic LTL expression} $\lstinline!;!$ )^{*} $\lstinline! }!\\%
\lstinline!    REQUIRE   { !%
$ ( \tlsfsec{basic LTL expression} $\lstinline!;!$ )^{*} $\lstinline! }!\\%
\lstinline!    ASSERT    { !%
$ ( \tlsfsec{basic LTL expression} $\lstinline!;!$ )^{*} $\lstinline! }!\\%
\lstinline!    ASSUME    { !%
$ ( \tlsfsec{basic LTL expression} $\lstinline!;!$ )^{*} $\lstinline! }!\\%
\lstinline!    GUARANTEE { !%
$ ( \tlsfsec{basic LTL expression} $\lstinline!;!$ )^{*} $\lstinline! }!\\%
\lstinline!  }!

\vspace{1em}

\noindent
All subsections except \lstinline!INPUTS! and \lstinline!OUTPUTS! are optional.

\subsection{Basic Expressions}
\label{sec:basicexpr}

A basic expression~$ e $ is either a boolean signal or a basic LTL
expression. Each basic expression has a corresponding type that is
$ \signals $ for boolean signals and $ \temporals $ for LTL
expressions.  Basic expressions can be composed to larger expressions
using operators.  An overview over the different types of expressions
and operators is given below.

\subsubsection{Boolean Signal Declarations}
A signal identifier is represented by a string consisting of lowercase
and uppercase letters~\mbox{('\lstinline!a!'-'\lstinline!z!'},
\mbox{'\lstinline!A!'-'\lstinline!Z!')},
numbers~\mbox{('\lstinline!0!'-'\lstinline!9!')},
underscores~\mbox{('\lstinline!_!')}, primes~\mbox{('\lstinline!'!')},
and at-signs~\mbox{('\lstinline!@!')} and does not start with a number
or a prime. Additionally, keywords like \lstinline!X!, \lstinline!G!
or \lstinline!U!, as defined in the rest of this document, are
forbidden.
An identifier is declared as either an input or an output signal. We
denote the set of declared input signals as $ \inputs $ and the set of
declared output signals as $ \outputs $, where
$ \inputs \cap \outputs = \emptyset $.  Then, a boolean signal
declaration simply consists of a signal identifier \tlsfid{name} from
$ \inputs \cup \outputs $.

\subsubsection{Basic LTL Expressions}
A basic LTL expression conforms to the following grammar, including
truth values, signals, boolean operators and temporal operators. For
easy parsing of the basic format, we require fully parenthesized
expressions, as expressed by the first of the following lines:
\begin{eqnarray*}
  \varphi & \equiv & 
  \text{\lstinline|(|} \varphi' \text{\lstinline|)|} \\
  \varphi' & \equiv & 
  \text{\lstinline|true|} \sep
  \text{\lstinline!false!} \sep
  s \text{\ \ \ for } s \in \inputs \cup \outputs \sep 
  \\ & &
  \text{\lstinline|!|} \varphi \sep
  \varphi \ \text{\lstinline!&&!} \ \varphi \sep
  \varphi \ \text{\lstinline!||!} \ \varphi \sep
  \varphi \ \text{\lstinline!->!} \ \varphi \sep
  \varphi \ \text{\lstinline!<->!} \ \varphi \sep
  \\ & & 
  \text{\lstinline!X!} \ \varphi \sep
  \text{\lstinline!G!} \ \varphi \sep
  \text{\lstinline!F!} \ \varphi \sep
  \varphi \ \text{\lstinline!U!} \ \varphi \sep
  \varphi \ \text{\lstinline!R!} \ \varphi \sep
  \varphi \ \text{\lstinline!W!} \ \varphi
\end{eqnarray*}
Thus, a basic LTL expression is either 
true, false, or a signal, or composed from these atomic expressions 
with boolean operators (negation,
conjunction, disjunction, implication, equivalence) and temporal
operators (next, globally, eventually, until, release, weak until).
The semantics of the boolean and temporal operators are defined in the usual
way.

\subsection{New Basic Expressions for \ltlf}
\label{sec:basicLTLf}

A basic \ltlf expression conforms to the following grammar. As before, 
we require fully parenthesized expressions in the basic format:
\begin{eqnarray*}
  \varphi & \equiv & 
  \text{\lstinline|(|} \varphi' \text{\lstinline|)|} \\
  \varphi' & \equiv & 
  \text{\lstinline|true|} \sep
  \text{\lstinline!false!} \sep
  s \text{\ \ \ for } s \in \inputs \cup \outputs \sep 
  \\ & &
  \text{\lstinline|!|} \varphi \sep
  \varphi \ \text{\lstinline!&&!} \ \varphi \sep
  \varphi \ \text{\lstinline!||!} \ \varphi \sep
  \varphi \ \text{\lstinline!->!} \ \varphi \sep
  \varphi \ \text{\lstinline!<->!} \ \varphi \sep
  \\ & & 
  \text{\lstinline!X!} \ \varphi \sep
  \text{\lstinline!G!} \ \varphi \sep
  \text{\lstinline!F!} \ \varphi \sep
  \varphi \ \text{\lstinline!U!} \ \varphi \sep
  \varphi \ \text{\lstinline!R!} \ \varphi \sep
  \varphi \ \text{\lstinline!W!} \ \varphi \sep
  \\ & & 
  \text{\lstinline|X[!]|} \ \varphi \sep
\end{eqnarray*}
Thus, a basic \ltlf expression can contain the same operators as an LTL expression, 
but additionally may contain the \emph{strong next} operator \text{\lstinline|X[!]|}.
Also note that \text{\lstinline|X|} is interpreted as \emph{weak next} in
\ltlf, as described in \Cref{sec:ltlf}.

\section{LTL over Finite Words}
\label{sec:ltlf}
We now recall the semantics of \ltlf over finite words as well as the
synthesis problem for \ltlf specifications.
As is clear from the previous section, the syntax of \ltlf is very similar to
that of LTL. In fact, syntactically it
only adds a \emph{strong next} operator.

For the rest of this section, we write $AP$ to denote the set $\inputs \cup
\outputs$ of all atomic
propositions.

\subsection{Semantics over Finite Words}
The semantics of an \ltlf formula $\varphi$ over finite words $\alpha =
a_0 a_1 \dots a_{n-1} \in (2^{AP})^*$ of lengths $n > 0$ are inductively defined
forall position $0 \le i < n$ as follows:
\begin{itemize}
  \item $\alpha, i \models \oper{true}$; $\alpha, i \not\models \oper{false}$; $\alpha, i \models p$ iff $p \in a_i$
  \item $\alpha, i \models \lnot \varphi$ iff $\alpha, i
    \not\models \varphi$
  \item $\alpha, i \models \varphi \ \oper{||} \ \psi$ iff
    $\alpha, i \models \varphi$ or $\alpha, i \models \psi$
  \item $\alpha, i \models \oper{X[!]} \varphi$ iff $i + 1 < n$ and
    $\alpha, i+1 \models \varphi$
  \item $\alpha, i \models \oper{X} \varphi$ iff $\alpha, i \models \lnot(\oper{X[!]}(\lnot \varphi))$
  
  \item $\alpha, i \models \oper{F} \varphi$ iff $\exists j$ s.t. $i\le j < n$ we have $\alpha, j \models \varphi$
  \item $\alpha, i \models \oper{G} \varphi$ iff $\forall j$ s.t. $i\le j < n$ we have $\alpha, j \models \varphi$
  \item $\alpha, i \models \varphi_1 \ \oper{U} \ \varphi_2$ iff $\exists
    k$ s.t. $i \leq k < n $ and $\alpha, k \models \varphi_2$;
    additionally, $\forall j$
    s.t. $\forall i \leq j < k$, we have $\alpha, j \models \varphi_1$
  \item $\alpha, i \models \varphi_1 \ \oper{W} \ \varphi_2$ iff
    $\varphi_1 \ \oper{U} \ \varphi_2$ or $\oper{G}
    \varphi_1$
\end{itemize}
Finally, we say $\alpha$ satisfies $\varphi$, written $\alpha \models
\varphi$, if $\alpha, 0 \models \varphi$ holds.


Note the asymmetry between the two versions of the next operator. The strong
version ($\oper{X[!]}\varphi$) requires that there be a next letter in the word
and to be compatible with $\varphi$, 
whereas for its weak counter part ($\oper{X}\varphi$) is also accepting 
if evaluated at the last position of the word.

\subsection{Synthesis and realizability}
There exist two possible interpretations for controllers obtained via synthesis from
\ltlf formulas as discussed in \cite{bansal2023model}:
terminating and non-terminating controllers.

The intuition behind it is that a terminating controller determines the length
of the finite word constructed.
The controller may, so to speak, explicitly state that the word constructed
so far satisfies the given formula and the word is now terminated.
A non-terminating controller on the other hand will always continue to produce
output signals, and such a controller implements the formula if
for every \textit{infinite} word constructed in this way there exists a \textit{finite}
prefix of it that satisfies the formula.

In addition to providing a clearer and more intuitive interpretation,
naturally we expect a controller to know when a task is finished. Importantly,
it is also known that
model checking terminating semantics has better computational
complexity~\cite{bansal2023model}. We therefore retain this choice for
\syntcomp.

\paragraph*{Definition}

Notice that there is a natural bijection between words in $(2^{AP})^*$ and those in
$(2^\inputs \times 2^\outputs)^*$ and $(2^\inputs)^* \times (2^\outputs)^*$.
For convenience, we work with words of the latter types while
still speaking of them as satisfying \ltlf formulas over $AP$ via said bijection.
A special (output) proposition, called \emph{alive signal} and denoted $as$, 
is added to the propositions of the formula in order to allow the controller 
to signal termination. 
This special proposition is an output signal as it is decided by the
controller, and it is guaranteed that $as$ will not appear in the atomic 
propositions used by the formula.
To avoid heavy notation we shall write $I = 2^\inputs$ and 
$O = 2^{\outputs \cup \left\lbrace as \right\rbrace}$.
%

As for LTL, \ltlf allows for Mealy and Moore semantics. In Mealy semantics,
the controller is a function
$f : I^* \times I \to O$ which can w.l.o.g. be represented by a terminating 
Mealy machine.
In Moore semantics, the output only depends on the current state so the
controller is a function 
$f : Q \to O$ which can w.l.o.g. be represented by a terminating 
Moore machine.

\paragraph*{Definition Terminating Mealy machine}

A Mealy machine is a tuple $M = (Q, \inputs, \outputs, q_0, T, \delta, \lambda)$
where $Q$ is a finite set of states; $\inputs$/$\outputs$ the set of atomic
input / output propositions; $q_0 \in Q$ is the initial state; $T \subset Q$ a set of terminating states;
$\delta: Q \times I \to Q$ the transition function and 
$\lambda: Q \times I \to O$ the output function.
To indicate termination, the alive signal is part of the output whenever the 
current state is not a terminating state: $\lambda(q,i) \models as$ if and only if
$q \not\in T$. Since the empty word is always rejected, $q_0$ must not be a
terminating state.

Transitions of such machines are depicted as $q \xrightarrow{i/o} q'$ 
with $\delta(q, i) = q'$ and $\lambda(q, i) = o$. 
We extend $\delta$ and $\lambda$ to words in $I^*$ in the usual inductive way

\begin{minipage}{0.49\linewidth}
  \begin{align*}
    & \delta^* : Q \times I^* \to Q \\
    & \delta^*(q, \varepsilon) = q \\
    & \delta^*(q, aw) =  \delta^*(\delta(q, a), w)
  \end{align*}
\end{minipage}
\hfill
\begin{minipage}{0.49\linewidth}
    \begin{align*}
    & \lambda^* : Q \times I^* \to O^* \\
    & \lambda^*(q, \varepsilon) = \varepsilon \\
    & \lambda^*(q, aw) =  \delta(q, a) \cdot \delta^*(q, w)
  \end{align*}
\end{minipage}

\noindent
with $a\in I$ and $w \in I^*$.

Note that, as usual, Mealy machines are deterministic
and need to be input complete. 
The language of the terminating Mealy machine is defined as
\[
  L_M = \left\lbrace \alpha \times \beta \mid \alpha \in I^*\colon\ \beta = \lambda^*(q_0, \alpha) 
\text{ and } \beta, \left| \beta \right| - 1  \not\models as \right\rbrace.
\]

Now, the Mealy machine $M$ is \emph{terminating} if 
for every infinite input sequence $\alpha_\omega \in I^\omega$ there exists a 
word $\alpha \times \beta$ in its language $L_M$ such that $\alpha$ is a prefix of 
$\alpha_\omega$.

We also define the \emph{reduced language} $L_{M,red}$ of $M$ as the subset of $L_M$
containing only the shortest terminating sequences. In symbols,
\[
  L_{M,red} = \left\lbrace \alpha \times b_0 \dots b_{n-1} \mid \alpha \in
    I^*\colon\ b_0 \dots b_{n-1} = \lambda^*(q_0, \alpha), as \not\in b_{n-1},
  \text{ and } b_0 \dots b_{n-2} \models \oper{G}\ as \right\rbrace.
\]

Finally we say that a terminating Mealy machine models a \ltlf formula 
$\varphi$ iff every word in its reduced language satisfies the formula.

To give a practical example, consider the formula
$\oper{G}(i_1 \leftrightarrow o_1) \land (o_2 \lor \oper{X[!]}\oper{true})$ 
with $\inputs = \left\lbrace i_1\right\rbrace$,
$\outputs = \left\lbrace o_1, o_2\right\rbrace$.
Two correct terminating Mealy machines are shown in fig.~\ref{figTermMealy}.


%
%
%

In practice, the alive signal is denoted by $\_\_AliveSig\_\_$ for all
benchmarks. 

\begin{figure}
  \centering
  \begin{tikzpicture}[automaton,node distance=2cm]
    \node[initial,state] (s0) {};
    \node[state, accepting,right=of s0] (s1) {};
    \path[->] (s0) edge[bend left] node[above=1mm]{$\left\lbrace i\right\rbrace/\left\lbrace o_1,o_2,as\right\rbrace$} (s1)
              (s0) edge[bend right] node[below=1mm]{$\left\lbrace \right\rbrace/\left\lbrace o_2,as\right\rbrace$} (s1) 
              (s1) edge[loop right] node[right=1mm, align=center]{$\left\lbrace i\right\rbrace/\left\lbrace o_1 \right\rbrace$\\$\left\lbrace \right\rbrace/\left\lbrace \right\rbrace$} (s1);

    \node[initial,state] (s00) at (0, -3){};
    \node[state,right=of s00] (s01) {};
    \node[state, accepting,right=of s01] (s02) {};
    \path[->] (s00) edge[bend left] node[above=1mm]{$\left\lbrace i \right\rbrace/\left\lbrace o_1,as \right\rbrace$} (s01)
              (s00) edge[bend right] node[below=1mm]{$\left\lbrace \right\rbrace/\left\lbrace as \right\rbrace$} (s01) 
              (s01) edge[bend left] node[above=1mm]{$\left\lbrace i \right\rbrace/\left\lbrace o_1,as \right\rbrace$} (s02)
              (s01) edge[bend right] node[below=1mm]{$\left\lbrace \right\rbrace/\left\lbrace as \right\rbrace$} (s02) 
              (s02) edge[loop right] node[right=1mm, align=center]{$\left\lbrace i \right\rbrace/\left\lbrace o_1 \right\rbrace$\\$\left\lbrace \right\rbrace/\left\lbrace \right\rbrace$} (s02);
  \end{tikzpicture}
  \caption{Two terminating Mealy machines satisfying 
  $\oper{G}(i_1 \leftrightarrow o_1) \land (o_2 \lor \oper{X[!]}\oper{true})$.
  The reduced language of the machine above consists of 
  $(\left\lbrace i \right\rbrace, \left\lbrace o_1,o_2,as \right\rbrace)$
  and 
  $(\left\lbrace \right\rbrace, \left\lbrace o_2,as \right\rbrace)$, both 
  satisfying $\oper{G}(i_1 \leftrightarrow o_1) \land o_2$.
  The reduced language of the machine below consists of 
  $(\left\lbrace i \right\rbrace, \left\lbrace o_1,as \right\rbrace)(\left\lbrace i \right\rbrace, \left\lbrace o_1,as \right\rbrace)$,
  $(\left\lbrace i \right\rbrace, \left\lbrace o_1,as \right\rbrace)(\left\lbrace \right\rbrace, \left\lbrace as \right\rbrace)$,
  $(\left\lbrace \right\rbrace, \left\lbrace as \right\rbrace)(\left\lbrace i \right\rbrace, \left\lbrace o_1,as \right\rbrace)$
  and 
  $(\left\lbrace \right\rbrace, \left\lbrace as \right\rbrace)(\left\lbrace \right\rbrace, \left\lbrace as \right\rbrace)$, all of them 
  satisfying $\oper{G}(i_1 \leftrightarrow o_1) \land \oper{X[!]}\oper{true}$.
  }
  \label{figTermMealy}
\end{figure}

\paragraph*{Definition Terminating Moore machine}

The definition of a terminating Moore machine and its language is very 
similar to the one of a terminating Mealy machine.
The only difference is that the output function no longer depends on the
(current) input: $\lambda: Q \to O$.
Note that the extension $\lambda^*$ of $\lambda$ to finite words does still depend
on the input. The only difference with respect to Mealy machines is that 
the last (output) letter of the word is removed.
All other definitions remain the same as for terminating Mealy machines.

The example formula used to showcase the terminating Mealy machines
$\oper{G}(i_1 \leftrightarrow o_1) \land (o_2 \lor \oper{X[!]}\oper{true})$
is not realizable in Moore semantics, as the output depends directly on the
input. We can however modify it and consider
$\oper{G}(i_1 \leftrightarrow \oper{X}(o_1)) \land (o_2 \lor \oper{X[!]}\oper{true})$
as shown in fig.~\ref{figTermMoore}.

\begin{figure}
  \centering
  \begin{tikzpicture}[automaton,node distance=1cm and 2cm]
    \node[initial,state] (s0) {};
    \node[state, accepting,right=of s0] (s1) {};
    \path[->] (s0) edge node[above=1mm, align=center]{$\left\lbrace i\right\rbrace/\left\lbrace o_2,as\right\rbrace$\\$\left\lbrace \right\rbrace/\left\lbrace o_2,as\right\rbrace$} (s1)
              (s1) edge[loop right] node[right=1mm, align=center]{$\left\lbrace i\right\rbrace/\left\lbrace \right\rbrace$\\$\left\lbrace \right\rbrace/\left\lbrace \right\rbrace$} (s1);

    \node[initial,state] (s00) at (0, -4){};
    \node[state, above right=of s00] (s01) {};
    \node[state, below right=of s00] (s02) {};
    \node[state, accepting,below right=of s01] (s03) {};
    \path[->] (s00) edge node[sloped, above=1mm]{$\left\lbrace i \right\rbrace/\left\lbrace as \right\rbrace$} (s01)
              (s00) edge node[sloped, below=1mm]{$\left\lbrace \right\rbrace/\left\lbrace as \right\rbrace$} (s02) 
              (s01) edge node[sloped, above=1mm, align=center]{$\left\lbrace i \right\rbrace/\left\lbrace o_1,as \right\rbrace$\\$\left\lbrace \right\rbrace/\left\lbrace o_1,as \right\rbrace$} (s03)
              (s02) edge node[sloped, below=1mm, align=center]{$\left\lbrace i \right\rbrace/\left\lbrace as \right\rbrace$ \\ $\left\lbrace \right\rbrace/\left\lbrace as \right\rbrace$} (s03) 
              (s03) edge[loop right] node[right=1mm, align=center]{$\left\lbrace i \right\rbrace/\left\lbrace \right\rbrace$\\$\left\lbrace \right\rbrace/\left\lbrace \right\rbrace$} (s03);
  \end{tikzpicture}
  \caption{Two terminating Moore machines satisfying 
  $\oper{G}(i_1 \leftrightarrow \oper{X}(o_1)) \land (o_2 \lor \oper{X[!]}\oper{true})$.
  The reduced language of the machine above consists of 
  $(\left\lbrace i \right\rbrace, \left\lbrace o_2,as \right\rbrace)$
  and 
  $(\left\lbrace \right\rbrace, \left\lbrace o_2,as \right\rbrace)$, both 
  satisfying $\oper{G}(i_1 \leftrightarrow \oper{X}(o_1)) \land o_2$ due to the 
  weak next operator.
  The reduced language of the machine below consists of 
  $(\left\lbrace i \right\rbrace, \left\lbrace as \right\rbrace)(\left\lbrace \right\rbrace, \left\lbrace o_1, as \right\rbrace)$,
  $(\left\lbrace i \right\rbrace, \left\lbrace as \right\rbrace)(\left\lbrace i \right\rbrace, \left\lbrace o_1, as \right\rbrace)$,
  $(\left\lbrace  \right\rbrace, \left\lbrace as \right\rbrace)(\left\lbrace \right\rbrace, \left\lbrace as \right\rbrace)$
  and
  $(\left\lbrace  \right\rbrace, \left\lbrace as \right\rbrace)(\left\lbrace i \right\rbrace, \left\lbrace as \right\rbrace)$
  all of them 
  satisfying $\oper{G}(i_1 \leftrightarrow \oper{X}(o_1)) \land \oper{X[!]}\oper{true}$.
  }
  \label{figTermMoore}
\end{figure}

\section{Targets and Semantics}
\label{sec:semantics}
\subsection{Targets}
The \lstinline!TARGET! of the specification defines the implementation model that
a solution should adhere to. Currently supported targets are Mealy
automata \mbox{(\lstinline!Mealy!)}, whose output depends on the
current state and input, and Moore automata
\mbox{(\lstinline!Moore!)}, whose output only depends on the current
state. The differentiation is necessary since realizability of a 
specification depends on the target system model. For example, every 
specification that is realizable
under Moore semantics is also realizable under Mealy semantics, but
not vice versa. 

\subsection{Semantics}
The \lstinline!SEMANTICS! of the specification defines how the formula
was intended to be evaluated, which also depends on an
implementation model.  We now support six different semantics,
which are standard Mealy semantics \mbox{(\lstinline!Mealy!)} and standard Moore
semantics \mbox{(\lstinline!Moore!)}, as well as their strict and finite variants
\mbox{(\lstinline!TARGET,!} \lstinline!VARIANT!), where $\text{\lstinline!VARIANT!} \in \{ \text{\lstinline!Strict!, \lstinline!Finite!} \}$. 

In the following, consider a specification where \lstinline!INITIALLY! evaluates to the LTL formula $\theta_e$, \lstinline!PRESET! evaluates to $\theta_s$, \lstinline!REQUIRE! evaluates to $\psi_e$, \lstinline!ASSERT! evaluates to $\psi_s$, \lstinline!ASSUME! evaluates to $\varphi_e$, and \lstinline!GUARANTEE! evaluates to $\varphi_s$. For specification sections that are not present, the respective formula is interpreted as \lstinline!true!.

\subsubsection{Standard semantics}
If the semantics is (standard) \mbox{\lstinline!Mealy!} or 
\mbox{\lstinline!Moore!}, and the \lstinline!TARGET! coincides with the semantics system 
model, then the specification is interpreted as the formula
\begin{equation}\label{eqn:std-semantics}
  \theta_e \rightarrow \left( \theta_s \land (\LTLglobally \psi_{e} \land \varphi_{e} \rightarrow \LTLglobally \psi_{s} \wedge \varphi_{s}) \right)
\end{equation}
in standard LTL semantics. 
Note that we require that the \lstinline!PRESET! property $\theta_s$ holds whenever the \lstinline!INITIALLY! condition $\theta_e$ holds, regardless of other environment assumptions.

\subsubsection{Strict semantics}
If the semantics is \mbox{\lstinline!Mealy,Strict!} or 
\mbox{\lstinline!Moore,Strict!}, and the \lstinline!TARGET! coincides with the semantics 
system model, then the specification is interpreted under strict implication 
semantics (as used in the synthesis of GR(1) specifications), which is equivalent to the formula
\begin{equation*}
  \theta_e \rightarrow \left( \theta_s \land (\psi_{s} \LTLweakuntil \neg \psi_{e}) \land (\LTLglobally \psi_{e} \land \varphi_{e} \rightarrow \varphi_{s}) \right)
\end{equation*}
in standard LTL semantics. In this case, we additionally require that the \lstinline!ASSERT! property $\psi_s$ needs to hold at least as long as the \lstinline!REQUIRE! condition $\psi_e$ holds. 

Note that this gives us an easy way to convert a specification with strict semantics into one with non-strict semantics. For details on strict implication semantics, see Klein and 
Pnueli~\cite{KleinP10}, as well as Bloem et al.~\cite{BloemJPPS12}, from which we also take our definition and interpretation of the GR(1) fragment.\footnote{Note that in the conversion of
  \cite{BloemJPPS12}, the formula is strengthened by
  adding the
  formula~$ \LTLglobally (\LTLpastglobally \psi_{e} \rightarrow
  \psi_{s})$,
  where $ \LTLpastglobally \varphi $ is a Past-LTL formula and denotes
  that $ \varphi $ holds everywhere in the past. However, it is easy
  to show that our definition of strict semantics matches the
  definition of \cite{BloemJPPS12}. We prefer this
  notion, since it avoids the introduction of Past-LTL.}

\subsubsection{Finite Semantics}
If the semantics is (finite) \mbox{\lstinline!Mealy,Finite!} or 
\mbox{\lstinline!Moore,Finite!}, and the \lstinline!TARGET! coincides with the semantics system 
model, then the specification is interpreted like in the standard semantics,
i.e. as the formula in Equation~\eqref{eqn:std-semantics}.
However, this time the formula is to be interpreted in the in the \ltlf
semantics (see \Cref{sec:ltlf}).


\section{The Full Format}
\label{sec:format}
In the full format, a specification consists of three sections: the
\lstinline!INFO! section, the \lstinline!GLOBAL! section and the \lstinline!MAIN! section. The \lstinline!GLOBAL!
section is optional.
\begin{equation*}
  \tlsfsec{info} [\tlsfsec{global}] \tlsfsec{main}
\end{equation*}
The \lstinline!INFO! section is the same as in the basic format, defined in
\secref{sec:basicinfo}. The \lstinline!GLOBAL! section can be used to define
parameters, and to bind identifiers to expressions that can be used
later in the specification. The \lstinline!MAIN! section is used as before, but
can use extended sets of declarations and expressions.

We define the \lstinline!GLOBAL! section in \secref{sec:global}, and the changes
to the \lstinline!MAIN! section compared to the basic format in
\secref{sec:main-full}. The extended set of expressions that can be
used in the full format is introduced in \secref{sec:expressions},
enumerations, extended signal and function declarations in
\secref{sec:enumerations} and \ref{sec:functions}, and additional
notation in \secref{sec:bigoperator}--\ref{sec:comments}.

\subsection{The GLOBAL Section}
\label{sec:global}

The \lstinline!GLOBAL! section consists of the \lstinline!PARAMETERS!
subsection, defining the identifiers that parameterize the specification, 
and the \lstinline!DEFINITIONS!
subsection, that allows to define functions, enumerations and to bind
identifiers to complex expressions. Multiple declarations need to be
separated by a '\lstinline!;!'. The section and its subsections are
optional.

\vspace{1em}

\noindent
\lstinline!  GLOBAL {!\\%
\lstinline!    PARAMETERS { !\\%
\lstinline!      !$ ( \tlsfid{identifier} $\lstinline! = !%
$ \tlsfsec{numerical expression} $\lstinline!;!$ )^{*} $\\%
\lstinline!    }!\\%
\lstinline!    DEFINITIONS  { !\\%
\lstinline!      !%
$ ((\tlsfid{function declaration} \sep \tlsfid{enum declaration} 
\sep \tlsfid{identifier} $%
\lstinline! = !$ \tlsfsec{expression}) $\lstinline!;!$ )^{*} $\\%
\lstinline!    }!\\%
\lstinline!  }!

\subsection{The MAIN Section}
\label{sec:main-full}

Like in the basic format, the \lstinline!MAIN! section contains the partitioning
of input and output signals, as well as the main
specification. However, signal declarations can now contain signal
buses, and LTL expressions can use parameters, functions, and
identifiers defined in the \lstinline!GLOBAL! section.

\vspace{1em}

\noindent
\lstinline!  MAIN {!\\%
\lstinline!    INPUTS    { !%
$ ( \tlsfsec{signal declaration} $\lstinline!;!$ )^{*} $\lstinline! }!\\%
\lstinline!    OUTPUTS   { !%
$ ( \tlsfsec{signal declaration} $\lstinline!;!$ )^{*} $\lstinline! }!\\%
\lstinline!    INITIALLY { !%
$ ( \tlsfsec{LTL expression} $\lstinline!;!$ )^{*} $\lstinline! }!\\%
\lstinline!    PRESET    { !%
$ ( \tlsfsec{LTL expression} $\lstinline!;!$ )^{*} $\lstinline! }!\\%
\lstinline!    REQUIRE   { !%
$ ( \tlsfsec{LTL expression} $\lstinline!;!$ )^{*} $\lstinline! }!\\%
\lstinline!    ASSERT    { !%
$ ( \tlsfsec{LTL expression} $\lstinline!;!$ )^{*} $\lstinline! }!\\%
\lstinline!    ASSUME    { !%
$ ( \tlsfsec{LTL expression} $\lstinline!;!$ )^{*} $\lstinline! }!\\%
\lstinline!    GUARANTEE { !%
$ ( \tlsfsec{LTL expression} $\lstinline!;!$ )^{*} $\lstinline! }!\\%
\lstinline!  }!

\vspace{1em}

\noindent
As before, all subsections except \lstinline!INPUTS! and \lstinline!OUTPUTS! are
optional.

\subsection{Expressions}
\label{sec:expressions}

An expression~$ e $ is either a boolean signal, an $ n $-ary signal
(called bus), an enumeration type, a numerical expression, a boolean expression, an LTL
expression, or a set expression. Each expression has a corresponding
type that is either one of the basic types:
$ \signals, \buses, \enums, \nats, \bools, \temporals $, or a recursively
defined set type~$ \mathcal{S}_{\arbitrary} $ for some type~$ \arbitrary $.

As before, an identifier is represented by a string consisting of
lowercase and uppercase
letters~(\mbox{'\lstinline!a!'-'\lstinline!z!'},
\mbox{'\lstinline!A!'-'\lstinline!Z!'}),
numbers~('\lstinline!0!'-'\lstinline!9!'),
underscores~('\lstinline!_!'), primes~('\lstinline!'!'), and
at-signs~('\lstinline!@!') and does not start with a number or a
prime.  In the full format, identifiers are bound to expressions of
different type. We denote the respective sets of identifiers by
$ \signalids $, $ \busids $, $ \enumids $, $ \natids $, $ \boolids $,
$ \temporalids $, and $ \atypeids $.  Finally, basic expressions can
be composed to larger expressions using operators. In the full format,
we do not require fully parenthesized expressions. If an expression is
not fully parenthesized, we use the precedence order given in
Table~\ref{tab:precedence}. An overview over the all types of
expressions and operators is given below.

\begin{table}
  \centering

  \renewcommand{\arraystretch}{1.05}

  \begin{tabular}{|c|l|l|c|c|}
    \hline 
    \textbf{Precedence} & \textbf{Operator} & \textbf{Description} & \textbf{Arity} &  \textbf{Associativity} \\
    \hline
    \hline
    \multirow{6}{*}{1} & \lstinline!+[!$ \cdot $\lstinline!]!\ (\lstinline!SUM[!$ \cdot $\lstinline!]!) & sum & \multirow{6}{*}{unary} & \\
      & \lstinline!*[!$ \cdot $\lstinline!]!\ (\lstinline!PROD[!$ \cdot $\lstinline!]!) & product & & \\
      & \lstinline!|!$ \cdots $\lstinline!|!\ (\lstinline!SIZE!) & size & & \\
      & \lstinline!MIN! & minimum & & \\
      & \lstinline!MAX! & maximum & & \\
      & \lstinline!SIZEOF! & size of a bus & & \\
    \hline
    2 & \lstinline!*!\ (\lstinline!MUL!) & multiplication & binary & left-to-right \\
    \hline
    \multirow{2}{*}{3} & \lstinline!/!\ (\lstinline!DIV!) & integer division & \multirow{2}{*}{binary} & \multirow{2}{*}{right-to-left} \\
      & \lstinline!%!\ (\lstinline!MOD!) & modulo & & \\
    \hline
    \multirow{2}{*}{4} & \lstinline!+!\ (\lstinline!PLUS!) & addition & \multirow{2}{*}{binary} & \multirow{2}{*}{left-to-right} \\
      & \lstinline!-!\ (\lstinline!MINUS!) & difference & & \\
    \hline 
    \multirow{2}{*}{5} & \lstinline!(*)[!$ \cdot $\lstinline!]!\ (\lstinline!CAP[!$ \cdot $\lstinline!]!) & intersection & \multirow{2}{*}{unary} & \\
      & \lstinline!(+)[!$ \cdot $\lstinline!]!\ (\lstinline!CUP[!$ \cdot $\lstinline!]!) & union &  & \\
    \hline
    6 & \lstinline!(\)!\ (\lstinline!(-)!,\lstinline!SETMINUS!) & set difference & binary & right-to-left \\
    \hline
    7 & \lstinline!(*)!\ (\lstinline!CAP!) & intersection & binary & left-to-right \\
    \hline
    8 & \lstinline!(+)! (\lstinline!CUP!) & union & binary & left-to-right \\
    \hline
    \multirow{6}{*}{9} & \lstinline!==!\ (\lstinline!EQ!) & equality & \multirow{6}{*}{binary} & \multirow{6}{*}{left-to-right} \\
      & \lstinline~!=~\ (\lstinline!/=!, \lstinline!NEQ!) & inequality & & \\
      & \lstinline!<!\ (\lstinline!LE!) & smaller than & & \\
      & \lstinline!<=!\ (\lstinline!LEQ!) & smaller or equal than & & \\
      & \lstinline!>!\ (\lstinline!GE!) & greater then & & \\
      & \lstinline!>=!\ (\lstinline!GEG!) & greater or equal than & & \\
    \hline
    10 & \lstinline!IN!\ (\lstinline!ELEM!, \lstinline!<-!) & membership & binary & left-to-right \\
    \hline
    \multirow{6}{*}{11} & \lstinline~!~\ (\lstinline!NOT!) & negation & \multirow{6}{*}{unary} & \\
       & \lstinline!X! & next & & \\
       & \lstinline|X[!]| & strong next & & \\
       & \lstinline!F! & finally & & \\
       & \lstinline!G! & globally & & \\
       & \lstinline!&&[!$ \cdot $\lstinline!]!\ (\lstinline!AND[!$ \cdot $\lstinline!]!, \lstinline!FORALL[!$ \cdot $\verb!]!) & conjunction & & \\
       & \lstinline!||[!$ \cdot $\lstinline!]!\ (\lstinline!OR[!$ \cdot $\lstinline!]!, \lstinline!EXISTS[!$ \cdot $\verb!]!) & disjunction & & \\
    \hline
    12 & \lstinline!&&!\ (\lstinline!AND!) & conjunction & binary & left-to-right \\
    \hline
    13 & \lstinline!||!\ (\lstinline!OR!) & disjunction & binary & left-to-right \\     
    \hline
    \multirow{2}{*}{14} & \lstinline!->!\ (\lstinline!IMPLIES!) & implication & \multirow{2}{*}{binary} & \multirow{2}{*}{right-to-left} \\   
       & \lstinline!<->!\ (\lstinline!EQUIV!) & equivalence &  & \\     
    \hline
    15 & \lstinline!W! & weak until & binary & right-to-left \\
    \hline
    16 & \lstinline!U! & until & binary & right-to-left \\
    \hline
    17 & \lstinline!R! & release & binary & left-to-right \\
    \hline
    18 & \!{\color{blue!70!black}\raisebox{-9pt}{\scalebox{1.7}{\textasciitilde}}} \vspace{-5pt} & pattern match & binary & left-to-right \\
    \hline
    19 & \lstinline!:! & guard & binary & left-to-right \\    
    \hline
  \end{tabular}

  \renewcommand{\arraystretch}{1}

  \caption{The table lists the precedence, arity and associativity of
    all expression operators. Also consider the alternative names in
    brackets which can be used instead of the symbolic
    representations.}
  \label{tab:precedence}
\end{table}

\subsubsection{Numerical Expressions}
A numerical expression~$ e_{\nats} $ conforms to the following
grammar:
\begin{eqnarray*}
  e_{\nats} & \equiv &
  i \text{\ \ \ for } i \in \natids \sep 
  n  \text{\ \ \ for } n \in \nats \sep
  e_{\nats} \ \text{\lstinline!+!} \ e_{\nats} \sep
  e_{\nats} \ \text{\lstinline!-!} \ e_{\nats} \sep
  e_{\nats} \ \text{\lstinline!*!} \ e_{\nats} \sep
  e_{\nats} \ \text{\lstinline!/!} \ e_{\nats} \sep
  e_{\nats} \ \, \text{\lstinline!\%!} \ \, e_{\nats} \\ 
  & &
  \text{\lstinline!|!} e_{\mathcal{S}_{\arbitrary}} \text{\lstinline!|!} \sep
  \text{\lstinline!MIN!} \ e_{\mathcal{S}_{\nats}} \sep
  \text{\lstinline!MAX!} \ e_{\mathcal{S}_{\nats}} \sep
  \text{\lstinline!SIZEOF!} \ s \ \ \text{ for } s \in \busids
\end{eqnarray*}
Thus, a numerical expression either represents an identifier (bound to
a numerical value), a numerical constant, an addition, a subtraction, a
multiplication, an integer division, a modulo operation, the size of a
set, the minimal/maximal value of a set of naturals, or the size (i.e., width) of a
bus, respectively. The semantics are defined in the usual way.

\subsubsection{Set Expressions}
A set expression~$ e_{\mathcal{S}_{\arbitrary}} $, containing elements
of type $ \mathbb{X} $, conforms to the following grammar:
\begin{eqnarray*}
  e_{\mathcal{S}_{\arbitrary}} & \equiv &
  i  \text{\ \ \ for } i \in \atypeids \sep
  \text{\lstinline!\{!} \, e_{\arbitrary} \text{\lstinline!,!} \, e_{\arbitrary}
  \text{\lstinline!,!} \ldots \text{\lstinline!,!} \, e_{\arbitrary} \,
  \text{\lstinline!\}!} \sep
  \text{\lstinline!\{!} \, e_{\nats} \text{\lstinline!,!} \, e_{\nats} \, 
  \text{\lstinline!..!}\,  e_{\nats} \, \text{\lstinline!\}!} \sep 
  \\ &&
  e_{\mathcal{S}_{\arbitrary}} \, \text{\lstinline!(+)!} \ e_{\mathcal{S}_{\arbitrary}} \sep
  e_{\mathcal{S}_{\arbitrary}} \, \text{\lstinline!(*)!} \ e_{\mathcal{S}_{\arbitrary}} \sep
  e_{\mathcal{S}_{\arbitrary}} \, \text{\lstinline!(\\)!} \ e_{\mathcal{S}_{\arbitrary}}
\end{eqnarray*}
Thus, the expression~$ e_{\mathcal{S}_{\arbitrary}} $ either
represents an identifier (bound to a set of values of type
$ \arbitrary $), an explicit list of elements of type $ \arbitrary $,
a list of elements specified by a range (for $ \arbitrary = \nats $),
a union of two sets, an intersection or a difference,
respectively. The semantics of a range expression
\lstinline!{!$ x $\lstinline!,!$ y $\lstinline!..!$ z $\lstinline!}!
are defined for $ x < y $ via:
\begin{equation*}
  \{ n \in \nats \mid x \leq n \leq z \wedge \exists j.\ n = x +
  j \cdot (y-x) \}.
\end{equation*}
The semantics of all other expressions are defined as usual. Sets
contain either positive integers, boolean expressions, LTL
expressions, buses, signals, or other sets of a specific type.

\subsubsection{Boolean Expressions}
A boolean expression~$ e_{\bools} $ conforms to the following
grammar:
\begin{eqnarray*}
  e_{\bools} & \equiv & 
  i  \text{\ \ \ for } i \in \boolids \sep
  e_{\arbitrary} \ \text{\lstinline!IN!} \ e_{\mathcal{S}_{\arbitrary}} \sep
  \text{\lstinline|true|} \sep
  \text{\lstinline!false!} \sep
  \text{\lstinline|!|}\,e_{\bools} \sep
  \\ & & 
  e_{\bools} \ \text{\lstinline!&&!} \ e_{\bools} \sep
  e_{\bools} \ \text{\lstinline!||!} \ e_{\bools} \sep
  e_{\bools} \ \text{\lstinline!->!} \ e_{\bools} \sep
  e_{\bools} \ \text{\lstinline!<->!} \ e_{\bools} \sep
  \\ & & 
  e_{\nats} \ \text{\lstinline!==!} \ e_{\nats} \sep
  e_{\nats} \ \text{\lstinline~!=~} \ e_{\nats} \sep
  e_{\nats} \ \text{\lstinline!<!} \ e_{\nats} \sep
  e_{\nats} \ \text{\lstinline!<=!} \ e_{\nats} \sep
  e_{\nats} \ \text{\lstinline!>!} \ e_{\nats} \sep
  e_{\nats} \ \text{\lstinline!>=!} \ e_{\nats} 
\end{eqnarray*}
Thus, a boolean expression either represents an identifier (bound to a
boolean value), a membership test, true, false, a negation, a
conjunction, a disjunction, an implication, an equivalence, or an
equation between two positive integers (equality, inequality, less
than, less or equal than, greater than, greater or equal than),
respectively. The semantics are defined in the usual way. Note that
signals are not allowed in a boolean expression, but only in an LTL
expression.

\subsubsection{LTL Expressions}
An LTL expression~$ \varphi $ conforms to the same grammar as a
boolean expression, except that it additionally includes signals and
temporal operators.
%
\begin{eqnarray*}
  \varphi & \equiv & \ldots \sep
  i \text{\ \ for } i \in \temporalids \sep
  s \text{\ \ for } s \in \signalids \sep
  b \text{\lstinline![!} e_{\nats} 
  \text{\lstinline!]!} \text{ for } b 
  \in \busids \sep
  \\ & & 
  b_{0} \ \, \text{\lstinline!==!} \ \, b_{1} \text{\ \ \ for } b_{j} \in \busids \text{ and } b_{1-j} \in \enumids \sep
  b_{0} \ \, \text{\lstinline~!=~} \ \, b_{1} \text{\ \ \ for } b_{j} \in \busids \text{ and } b_{1-j} \in \enumids \sep
  \\ & &
  \text{\lstinline!X!} \ \varphi \sep
  \text{\lstinline|X[!]|} \ \varphi \sep
  \text{\lstinline!G!} \ \varphi \sep
  \text{\lstinline!F!} \ \varphi \sep
  \varphi \ \text{\lstinline!U!} \ \varphi \sep
  \varphi \ \text{\lstinline!R!} \ \varphi \sep
  \varphi \ \text{\lstinline!W!} \ \varphi 
\end{eqnarray*}
\goodbreak
\noindent
Thus, an LTL expression additionally can represent an identifier bound
to an LTL formula, a signal, an $e_{\nats} $-th signal of a bus, a
next operation, a restriction of a bus to a set of enumeration
valuations via equality or inequality, a globally operation, an
eventually operation, an until operation, a release operation, or a
weak until operation, respectively. Note that every boolean expression
is also an LTL expression, thus we allow the use of identifiers that
are bound to boolean expressions as well. 
The semantics of expressions involving bus
operations is defined in the subsequent sections.

\subsection{Enumerations}
\label{sec:enumerations}

An enumeration declaration conforms to the following grammar:
\begin{equation*}
  \text{\lstinline!enum!} \ \tlsfid{enumtype} \ \text{\lstinline!=!} \ 
  \Big( \tlsfid{identifier} \ \text{\lstinline!:!} \ 
  (\text{\lstinline!0!} \!\! \sep \!\! \text{\lstinline!1!} \!\! \sep 
  \!\! \text{\lstinline!*!})^{n} \big( \text{\lstinline!,!} \ 
  (\text{\lstinline!0!} \!\! \sep \!\! \text{\lstinline!1!} \!\! \sep
  \!\! \text{\lstinline!*!})^{n} \big)^{*} \Big)^{+}
\end{equation*}
for some arbitrary but fix positive integer $ n > 0 $. As an example consider the
enumeration \lstinline!Positions!, which declares the enumeration
identifiers \lstinline!LEFT!, \lstinline!MIDDLE!, \lstinline!RIGHT!, and
\lstinline!UNDEF! as members of $ \enumids $:

\vspace{0.6em}

\noindent
\lstinline!  enum Position =!\\%
\lstinline!    LEFT:   100!\\%
\lstinline!    MIDDLE: 010!\\%
\lstinline!    RIGHT:  001!\\%
\lstinline!    UNDEF:  11*, 1*1, *11!\\
   
\vspace{0.6em}

\noindent We use \lstinline!0! to identify the absent signal,
\lstinline!1! to identify the present signal and \lstinline!*! for
either of both. Each identifier then refers to at least one concrete
signal valuation sequence. Multiple values can be denoted by sequences
with a \lstinline!*!, as well as by comma separated
lists. Furthermore, the identifier of each declared valuation has to
be unique. Not all possible valuations have to be identified.

Enumeration identifiers can only be used in comparisons against buses
inside an LTL expression, where we require that the corresponding bus
has the same width as the valuation compared to. It defines a boolean
constraint on the bus, restricting it to the different valuations,
bound to the identifier, e.g., the expressions %
\lstinline!b == RIGHT! and %
\lstinline~!b[0] && !b[1] && b[2]~ are semantically equivalent, as
well as %
\lstinline!b /= UNDEF! and %
\lstinline~!((b[0] && b[1]) || (b[0] && b[2]) || (b[1] && b[2]))~.

\subsection{Signals and Buses}
\label{sec:signals}

A single signal declaration consists of the name of the signal.  As
for the basic format, signals are declared as either input or output
signals, denoted by $ \inputs $ and $ \outputs $, respectively. A bus
declaration additionally specifies a signal width, i.e., a bus
represents a finite set of signals. The signal width is either given
by a numerical value or via an enumeration type.
\begin{equation*}
  \tlsfid{name} \sep
  \tlsfid{name} \text{\lstinline![!} 
  e_{\nats} \text{\lstinline!]!} \sep 
  \tlsfid{enumtype} \tlsfid{name}
\end{equation*}
Semantically, a signal declaration \lstinline!s! specifies a signal
$ s \in \inputs \cup \outputs $, where a bus declaration
\lstinline!b[n]! specifies $ n $ signals~\lstinline!b[0]!,
\lstinline!b[1]!, $ \ldots $, \lstinline!b[n-1]!, with either
\lstinline!b[i]!$ \in \inputs $ for all $ 0 \leq i < n $, or
\lstinline!b[i]!$ \in \outputs $ for all $ 0 \leq i < n $. A bus
specified via an enumeration type has the same width as the valuations
of the corresponding enumeration.

Buses which are declared using an enumeration type, where not all
valuations are related to an identifier\footnote{See e.g.\ the
  \lstinline!000!  valuation of the example of \secref{sec:signals}}
induce an implicit constraint on the corresponding signals: if
the bus corresponds to a set of input signals, then the global
requirement that no other than the defined valuations appear on this
bus is imposed. If it corresponds to a set of output signals, then the
equivalent global invariant is imposed.

Finally, note that we use \lstinline!b[i]! to access the $ i $-th
value of $ b $, i.e., we use the same syntax as for the declaration
itself\hspace{1pt}\footnote{C-Array Syntax Style}. Also note that
for the declared signals~$ s $, we have
$ s \in \inputs \cup \outputs \subseteq \signalids $, and for the
declared buses $ b $, we have $ b \in \busids $.

\subsection{Function Declarations}
\label{sec:functions}

As another feature, one can declare (recursive) functions of arbitrary
arity inside the \lstinline!DEFINITIONS! section. Functions can be used to define
simple macros, but also to generate complex formulas from a given set
of parameters. A declaration of a function of arity~$ n $ has the form
%
\begin{equation*}
  \tlsfid{function name} \text{\lstinline!(!}
  \tlsfid{arg\ensuremath{_{1}}} \text{\lstinline!,!} 
  \tlsfid{arg\ensuremath{_{2}}} \text{\lstinline!,!}  
  \ldots \text{\lstinline!,!}
  \tlsfid{arg\ensuremath{_{n}}} \text{\lstinline!) =\ !}
  (e_{c})^{+},
\end{equation*}
where
$ \tlsfid{arg\ensuremath{_{1}}}, \tlsfid{arg\ensuremath{_{2}}},
\ldots, \tlsfid{arg\ensuremath{_{n}}} $
are fresh identifiers that can only be used inside the
sub-expressions~$ e_{c} $. An expression~$ e_{c} $ conforms to the
following grammar:
\begin{equation*}
  e_{c} \ \, \equiv \ \, e \sep
  e_{\bools} \ \text{\lstinline!:!} \ e \sep
  e_{\mathbb{P}} \ \text{\lstinline!:!} \ e 
  \qquad \qquad \text{where } 
  \ \ e \ \, \equiv  \ \,
  e_{\nats} \sep
  e_{\bools} \sep
  e_{\mathcal{S}_{\arbitrary}} \sep
  \varphi
\end{equation*}
Thus, a function can be bound to any expression~$ e $, parameterized
in its arguments, which additionally may be guarded by some boolean
expression~$ e_{\bools} $, or a pattern match~$ e_{\pats} $. If the
regular expression~$ (e_{c})^{+} $ consists of more than one
expression~$ e_{c} $, then the function binds to the first expression
whose guard evaluates to \lstinline!true! (in the order of their
declaration). Furthermore, the special guard \lstinline!otherwise! can
be used, which evaluates to \lstinline!true! if and only if all other
guards evaluate to \lstinline!false!. Expressions without a guard are
implicitly guarded by \lstinline!true!. All sub-expressions~$ e_{c} $
need to have the same type $ \arbitrary $. For every instantiation of
a function by given parameters, we view the resulting expression
$ e_{\arbitrary} $ as an identifier in~$ \Gamma_{\arbitrary} $, bound
to the result of the function application.
%
 
\subsubsection{Pattern Matching}
\label{sec_patterns}
Pattern matches are special guards of the form
\begin{equation*}
  e_{\pats} \equiv \ \, \varphi \ \lstilde \ \varphi' ,
  \vspace{-0.5em}
\end{equation*}
which can be used to describe different behavior depending on the
structure of an LTL expression. Hence, a guard~$ e_{\pats} $ evaluates
to \lstinline!true! if and only if $ \varphi $ and $ \varphi' $ are
structurally equivalent, with respect to their boolean and temporal
connectives. Furthermore, identifier names that are used in
$ \varphi' $ need to be fresh, since every identifier expression that
appears in $ \varphi' $ is bound to the equivalent sub-expression in
$ \varphi $, which is only visible inside the right-hand-side of the
guard. Furthermore, to improve readability, the special
identifier~\lstinline!_! (wildcard) can be used,
which always remains unbound. To clarify this feature, consider the
following function declaration:

\vspace{0.6em}

\noindent
\lstinline!  fun(f) =!\\%
\vspace{-0.4em}%
\lstinline!    f !$ \hspace{-1.5pt} \lstilde \hspace{-1.5pt} $%
\lstinline! a U _: a!\\%
\lstinline!    otherwise: X f!
   
\vspace{0.6em}

\noindent The function $ \textit{fun} $ gets an LTL formula $ f $ as a
parameter. If $ f $ is an until formula of the form
$ \varphi_{1} \LTLuntil \varphi_{2} $, then $ \textit{fun}(f) $ binds
to $ \varphi_{1} $, otherwise $ \textit{fun}(f) $ binds to
$ \LTLnext f $.

\subsection{Big Operator Notation}
\label{sec:bigoperator}

It is often useful to express parameterized expressions using ``big''
operators, e.g., we use $ \Sigma $ to denote a sum over multiple
sub-expressions, $ \Pi $ to denote a product, or $ \bigcup $ to denote
a union. It is also possible to use this kind of notion in this
specification format. The corresponding syntax looks as follows:
\begin{equation*}
  \tlsfid{op} \text{\lstinline![!} \, 
  \tlsfid{id\ensuremath{_{0}}} \, 
  \text{\lstinline!IN!} \; e_{\mathcal{S}_{\arbitrary_{0}}} \! 
  \text{\lstinline!,!}\, \tlsfid{id\ensuremath{_{1}}} \, 
  \text{\lstinline!IN!} \; e_{\mathcal{S}_{\arbitrary_{1}}} \! 
  \text{\lstinline!,!}\, \ldots \, 
  \text{\lstinline!,!} \,\tlsfid{id\ensuremath{_{n}}} \, 
  \text{\lstinline!IN!} \; e_{\mathcal{S}_{\arbitrary_{n}}} 
  \text{\lstinline!]!} \, e_{\arbitrary} 
\end{equation*}
Let $ x_{j} $ be the identifier represented by
$ \tlsfid{id\ensuremath{_{j}}} $ and $ S_{j} $ be the set represented by
$ e_{\mathcal{S}_{\arbitrary_{j}}}\! $. Further, let
$ \bigoplus $ be the mathematical operator corresponding to
$ \tlsfid{op} $. Then, the above expression corresponds to the
mathematical expression:
\begin{equation*}
  \bigoplus\limits_{x_{0} \in S_{0}} \ 
  \bigoplus\limits_{x_{1} \in S_{1}} \ \cdots \ 
  \bigoplus\limits_{x_{n} \in S_{n}}
  \big( e_{\arbitrary} )
\end{equation*}
Note that $ \tlsfid{id\ensuremath{_{0}}} $ is already bound in
expression $ e_{\mathcal{S}_{\arbitrary_{1}}} \!$,
$ \tlsfid{id\ensuremath{_{1}}} $ is bound in
$ e_{\mathcal{S}_{\arbitrary_{2}}} \!$, and so forth. The syntax is
supported by every operator
$ \tlsfid{op} \in \{ \text{\lstinline!+!},
\text{\lstinline!*!}, \text{\lstinline!(+)!},
\text{\lstinline!(*)!}, \text{\lstinline!&&!},
\text{\lstinline!||!} \} $.

\subsection{Syntactic Sugar}
\label{sec:syntacticsugar}

To improve readability, there is additional syntactic sugar, which can
be used beside the standard syntax. Let $ n $ and $ m $ be numerical
expressions, then

\begin{itemize}

\item \lstinline!X[!$ n $\lstinline!]!$ \; \varphi $ denotes a stack
  of $ n $ next operations, e.g.: \\[0.5em]
  \lstinline!  X[3] a!$ \ \, \equiv \ \, $ \lstinline!X X X a!

\item \lstinline!F[!$ n $\lstinline!:!$ m $\lstinline!]!$ \; \varphi $
  denotes that $ \varphi $ holds somewhere between the next $ n $ and
  $ m $ steps, e.g.: \\[0.5em]
  \lstinline!  F[2:3] a!$ \ \, \equiv \ \, $\lstinline!X X(a || X a)!

\item \lstinline!G[!$ n $\lstinline!:!$ m $\lstinline!]!$ \; \varphi $
  denotes that $ \varphi $ holds everywhere between the next $ n $ and
  $ m $ steps, e.g.: \\[0.5em]
  \lstinline!  G[1:3] a!$ \ \, \equiv \ \, $\lstinline!X(a && X(a && X a))!

\item All of the above have been overloaded so that, with the finite
  semantics, they use the strong next operator if a \lstinline|!| appears
  inside (i.e. at the beginning or the end) the square braces. For instance: \\[0.5em]
  \lstinline|  G[!1:3] a|$ \ \, \equiv \ \,$\lstinline|G[1:3!] a|$ \ \, \equiv \ \, $\lstinline|X[!](a && X[!](a && X[!] a))|

\item $ \tlsfid{op} $\lstinline![!$ \, \ldots $\lstinline!,!$ \, n \, \circ_{1} 
  \tlsfid{id} \circ_{2} \, m \, $\lstinline!,!$ \ldots $\lstinline!]!$ \, e_{X} $
  denotes a big operator application, where  $ n \, \circ_{1} \tlsfid{id} 
  \circ_{2} \, m $ with $ \circ_{1}, \circ_{2} \in \{ \text{\lstinline!<!},
  \text{\lstinline!<=!} \} $ denotes that \tlsfid{id} ranges from $ n $ to
  $ m $. The inclusion of $ n $ and $ m $ depends on the
  choice of $ \circ_{1} $ and $ \circ_{2} $, respectively. Thus, the
  notation provides an alternative to membership in combination with
  set ranges, e.g.: \\[0.5em]
  \lstinline!  &&[0 <= i < n] b[i]!$ \ \, \equiv \ \, 
  $\lstinline!&&[i IN {0,1..n-1}] b[i]!
  %

\end{itemize}

\subsection{Comments}
\label{sec:comments}

It is possible to use C style comments anywhere in the specification,
i.e., there are single line comments initialized by \lstinline!//! and
multi line comments between \lstinline!/*! and
\lstinline!*/!.  Multi line comments can be nested.

%
\section{Updates to the SyFCo Tool}
\label{sec:tool}
We have updated the Synthesis Format Conversion Tool\footnote{See
\url{https://github.com/reactive-systems/syfco}} (SyFCo) to support TLSF
specifications over finite words. For this, we now support output in the
format \verb|ltlxba-fin|. The output format is similar to versions of \ltlf
that the Spot tool~\cite{duret.22.cav}
can handle.

%

\section*{Acknowledgements}
We would like to thank Luca Geatti, Nicola Gigante, Antonio Di Stasio, and
Shufang Zhu for their valuable feedback on earlier versions of this note. We
also thank Alexandre Duret-Lutz for pushing us to further formalize the
expectations in terms of controllers synthesized for the \ltlf track.

\bibliographystyle{eptcs}
\bibliography{biblio,synthesis}
%
%

\end{document}